\documentclass[useAMS,usenatbib,usegraphicx]{mn2e}

\usepackage{subfigure}
\usepackage{url}
\usepackage{comment}
\usepackage{color}
\usepackage{dcolumn}
\usepackage{enumitem}
\usepackage{hyperref}
\usepackage{longtable}
\usepackage{amssymb}
\usepackage[document]{ragged2e}
\usepackage{wrapfig}
\usepackage{hanging}
\usepackage[table]{xcolor}
\usepackage{colortbl}
\usepackage{hhline}
\usepackage{amsmath}

\newcommand*{\rom}[1]{\uppercase\expandafter{\romannumeral #1}}
\newcommand\ion[2]{#1$\;${\small\rmfamily\rom{#2}}\relax}%
\newcommand{\omgCen}{$\omega$ Cen }
\newcommand{\omgCennosp}{$\omega$ Cen}
\newcommand{\myA}{\AA \hspace{0.015in}}
\newcommand{\myAA}{\AA \hspace{0.025in}}
\newcommand{\cms}[1]{{\color{black}{#1}}}

\newcommand\aap{{A\&A}}%
\newcommand\araa{{ARA\&A}}%
\newcommand\aapr{A\&ARv}%
\newcommand\mnras{{MNRAS}}%
\newcommand\apj{{ApJ}}%
\newcommand\apjs{{ApJS}}%
\newcommand\apjl{{ApJ}}%
\newcommand\aj{{AJ}}%
\newcommand\pasp{{PASP}}%
\newcommand\memsai{{Mem.~Soc.~Astron.~Italiana}}%
\newcommand\procspie{{Proc.~SPIE}}%
\newcommand\nat{{\it Nature}}%

\setlist[enumerate]{noitemsep}
\setlist[enumerate,1]{leftmargin=*}
\setlist[itemize]{noitemsep}
\setlist[itemize,1]{leftmargin=*}
\setlist[description]{noitemsep}
\setlist[description,1]{leftmargin=*}
\setenumerate[0]{label=(\arabic*)}

\makeatother

\title[G1 Detailed Abundances]{The Massive M31 Cluster G1: Detailed
  Chemical Abundances from Integrated Light Spectroscopy\thanks{Based on observations obtained with the Hobby-Eberly Telescope, which is a joint project of the University of Texas at Austin, the Pennsylvania State University, Ludwig-Maximilians-Universit\"{a}t M\"{u}nchen, and Georg-August-Universit\"{a}t G\"{o}ttingen.}}

\author[Sakari et al.]{Charli M. Sakari$^{1}$\thanks{E-mail:
    sakaricm@sfsu.edu}, Matthew D. Shetrone$^{2,3}$, Andrew McWilliam$^{4}$,
  George Wallerstein$^{5}$\\
$^{1}$ San Francisco State University, 1600 Holloway Avenue, San Francisco, CA, 94132 USA\\
$^{2}$ University of California Observatories/Lick Observatory, University of California, Santa Cruz, CA, 95064, USA\\
$^{3}$ McDonald Observatory, University of Texas at Austin, HC75 Box
  1337-MCD, Fort Davis, TX 79734, USA\\
$^{4}$ Observatories of the Carnegie Institute of Washington,
Pasadena, CA, USA\\
$^{5}$ Department of Astronomy, University of Washington, Seattle WA
98195-1580, USA\\
}

\begin{document}

\maketitle

\label{firstpage}

\begin{abstract}
G1, also known as Mayall~II, is one of the most massive star clusters
in M31.  Its mass, ellipticity, and location in the outer halo make it
a compelling candidate for a former nuclear star cluster. This paper
presents an integrated light abundance analysis of G1, based on a
moderately high-resolution ($R=15,000$) spectrum obtained with the
High Resolution Spectrograph on the Hobby-Eberly Telescope in 2007 and
2008.   To independently determine the metallicity, a moderate
resolution ($R\sim4,000$) spectrum of the calcium-II triplet lines in
the near-infrared was also obtained with the Astrophysical Research
Consortium's 3.5-m telescope at Apache Point Observatory. From the
high-resolution spectrum, G1 is found to be a moderately metal-poor
cluster, with $[\rm{Fe/H}]~=~-0.98\pm0.05$.  G1 also shows signs of
$\alpha$-enhancement (based on Mg, Ca, and Ti) and lacks the
$s$-process enhancements seen in dwarf galaxies (based on comparisons
of Y, Ba, and Eu), indicating that it originated in a fairly massive
galaxy. Intriguingly, G1 also exhibits signs of Na and Al enhancement,
a unique signature of GCs---this suggests that G1's formation is
intimately connected with GC formation.  G1's high [Na/Fe] also
extends previous trends with cluster velocity dispersion to an even
higher mass regime, implying that higher mass clusters are more able
to retain Na-enhanced ejecta.  The effects of intracluster abundance
spreads are discussed in a subsequent paper.  Ultimately, G1's
chemical properties are found to resemble other M31 GCs, though it
also shares some similarities with extragalactic nuclear star
clusters.
\end{abstract}


\section{Introduction}\label{sec:Intro}
Historically, globular clusters (GCs) were once regarded as simple
stellar populations, i.e., dark-matter-free, chemically-homogeneous
spheroids that were fundamentally distinct from the more chemically
complex galaxies.  It is now widely accepted, however, that GCs are
not simple: all GCs in the Milky Way and its satellites show some
evidence for star-to-star chemical inhomogeneity, from variations in
carbon and nitrogen (seen in all massive GCs, including the
intermediate-age LMC GCs; e.g.,
\citealt{Hollyhead2017,Hollyhead2019}), to variations in sodium and
oxygen (seen in all classical, old GCs; e.g., \citealt{Carretta2009}),
to variations in heavier elements in a handful of GCs, including
neutron-capture elements \citep{Sneden1997,Roederer2011} and even
iron.  Iron is one of the more puzzling elements that can vary within
GCs, as iron spreads are generally interpreted as a signature of
multiple bursts of star formation.  However, some of the most massive
GCs exhibit clear spreads in [Fe/H], including the most massive Milky
Way (MW) GC, \omgCen, which shows at least four discrete populations
with an iron spread of $\sim 2$ dex
\citep{FreemanRogers1975,JohnsonPilachowski2010,Villanova2014}.  There
are at least five more of these ``iron-complex'' GCs in the MW with
clear, significant iron spreads
\citep{Carretta2010,Marino2011,Marino2015,Yong2014,Johnson2017}.
These iron-complex GCs also host Na/O variations, just like the
classical GCs; it is uncertain how these iron-complex GCs fit in with
the general GC population.

A potential explanation for the iron-complex GCs is that they were
once the nuclear star clusters (NSCs) of dwarf galaxies which have
since been accreted into the MW.   In the NSC framework, the iron
spreads are then caused by ongoing {\it in situ} star formation,
mergers of multiple classical GCs, or some mixture of the two
scenarios (see the review by \citealt{Neumayer2020}).  There is some
observational evidence to suggest that the iron-complex GCs did
originate in dwarf galaxies. One of the iron-complex GCs, M54
\citep{Carretta2010}, lies very close to the expected core of the
Sagittarius dwarf spheroidal, a galaxy which is actively being
accreted into the MW halo \citep{Ibata1995}.  \citet{Johnson2017} also
found several stars in M19 with low [$\alpha$/Fe], a common signature
of metal-rich dwarf galaxy stars (e.g., \citealt{Tolstoy2009}).
Most of the iron-complex GCs in the MW, however, are dominated by
metal-poor, $\alpha$-enhanced stars that follow the MW's standard
chemical evolution track (e.g., \citealt{Marino2015,Johnson2017}),
\cms{making it difficult to identify accreted GCs through chemical
abundance analyses.} Other groups have identified tidal streams or field star
populations around the iron-complex GCs (e.g., \citealt{Ibata2019}),
and several iron-complex GCs have been linked to streams and dwarf
galaxies, including Gaia-Enceladus
\citep{Helmi2018,Massari2019}. Unraveling the mystery of the
iron-complex GCs requires identifying more clusters and studying their
detailed chemical abundances and kinematics.

Massive GCs are not unique to the MW; other galaxies have GCs
that are even more massive than \omgCen, including M31.  One such
cluster is G1 (also known as Mayall II; \citealt{MayallEggen1953}).
G1's total mass has been estimated to be $\sim
7-17\times10^{6}\;\rm{M}_{\sun}$, making it at least twice as massive
as \omgCen \citep{Meylan2001}.  Photometric evidence suggests that G1 is
also an iron-complex GC: the width of its red giant branch (RGB) in an
{\it Hubble Space Telescope (HST)} CMD indicates an [Fe/H]
dispersion of $0.1-0.5$ dex
\citep{Meylan2001,Nardiello2019}.\footnote{\citet{Meylan2001}
  estimated $\Delta[\rm{Fe/H}]=0.4-0.5$ dex, but \citet{Nardiello2019}
  have subsequently revised this estimate down to $\Delta[\rm{Fe/H}]
  \sim 0.15$ dex.}  G1 has also gained some notoriety for being the
potential host of an intermediate-mass black hole
\citep{Gebhardt2002,Gebhardt2005}, though \citealt{MillerJones2012}
argue that the X-ray and radio data are more indicative of a low-mass
X-ray binary.  \citet{Baumgardt2003} argue that G1's dynamics match
models for a GC-GC merger, which would be consistent with an origin as
a NSC.

G1 also has indications that it may have originated in a dwarf
galaxy.  It seems to be a relatively metal-rich GC, at $[\rm{Fe/H}]
\sim -0.7$ to $-1$ \citep{Rich1996,Meylan2001,Perina2012} despite its
location in the outer halo ($R_{\rm{proj}}=34.7$ kpc from the
centre of M31; \citealt{Mackey2019}). Simulations by
\citet{BekkiChiba2004} confirm that G1 could indeed have been created
by stripping a nucleated dwarf of its outer envelope of stars.
However, tidal debris has not been found around G1
\citep{Reitzel2004}.  \citet{Mackey2010} also noted that G1 appears to
lie within a group of GCs on the western side of M31, known as
``Association 2,'' but \citet{Veljanoski2014} found that G1's
radial velocity did not match the other GCs.  G1's presence in the
outer halo is therefore somewhat of a mystery; there are strong
indications that it may have been a NSC, yet there is no sign of the
rest of the galaxy.  There are also other similarly massive clusters
in M31, though none are obviously in the outer halo (but
see \citealt{Perina2012}).  Detailed chemical abundances of a wide
variety of elements from stars in this moderately metal-poor cluster
can shed light on its possible status as a GC and as a former dwarf
galaxy NSC.  However, G1's distance renders its brightest stars
too faint for high resolution, high S/N spectroscopy, since the
brightest stars have $V\sim 22$ mag.

Fortunately, G1 can be studied through high resolution integrated
light (IL) spectroscopy, where a single spectrum is obtained from an entire
stellar population.  The capabilities and limitations of
high-resolution IL spectroscopy have been thoroughly laid out by
\citet{McWB}, \citet{Colucci2009,Colucci2011,Colucci2012,Colucci2014},
and \citet{Sakari2013,Sakari2014,Sakari2015,Sakari2016}.  Briefly,
high resolution IL spectroscopy can produce flux-weighted average
abundances of many elements, including Fe, Mg, Ca, Ba, and
Eu. \cms{Tests with Milky Way GCs have demonstrated that IL abundances
  match the values from individual stars:} when the elements do not
vary within a cluster, these integrated abundances represent the
primordial values; otherwise, the integrated abundances fall within
the observed ranges (e.g., \citealt{McWB,Sakari2013,Colucci2017}).
Though G1's putative Fe spread complicates these analyses, the
uncertainties can be robustly quantified, and the possibilities for
identifying an Fe spread can be explored---this will be the subject of
a forthcoming paper.

This paper therefore presents the first detailed chemical abundances
in the massive M31 GC, G1.  A medium resolution calcium II triplet
(CaT) spectrum is also presented as an independent verification of the
average cluster metallicity.  Section \ref{sec:Observations} presents
the observations and data reduction, while Section \ref{sec:Isochrone}
discusses the identification of an appropriate isochrone to model the
underlying stellar population and the resulting chemical
abundances. G1's status as a GC, as a potential NSC, and as a member
of M31's outer halo are then discussed in Section
\ref{sec:Discussion}.  Given the likelihood of intra-cluster abundance
spreads within G1, the systematic offsets resulting from undetected
abundance spreads are quantified in a forthcoming paper (Sakari et
al., {\it in prep.}).

\clearpage

\section{Observations and Data Reduction}\label{sec:Observations}

\begin{figure*}
\begin{center}
\centering
\hspace*{-0.25in}
\includegraphics[scale=0.65,trim=1.2in 0 1.0in 0.0in,clip]{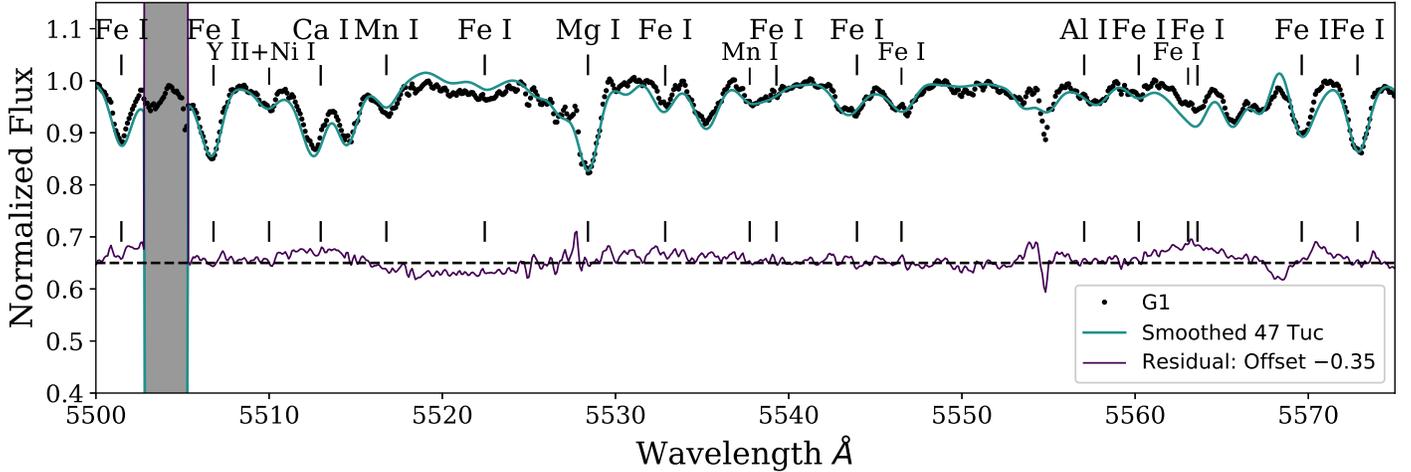}
\caption{A comparison between the optical, high-resolution IL spectra
  of G1 (points) and 47 Tuc (the thick green line; from
  \citealt{McWB}), where the 47~Tuc spectrum has been smoothed to the
  velocity dispersion of G1.  The thin purple line (arbitrarily offset
  below the two spectra) shows the residuals.  The grey region shows a
  bad region of the 47~Tuc spectrum.}\label{fig:47TucComp1}
\end{center}
\end{figure*}

\subsection{High Resolution Spectrum}\label{subsec:HRObservations}
The G1 data were obtained at McDonald Observatory in Fort Davis, TX
using the High Resolution Spectrograph (HRS;  \citealt{HRSref}) on the
Hobby-Eberly Telescope (HET; \citealt{HETref,HETQueueref}).
Observations were carried out in 2007 and 2008 (see Table
\ref{table:Targets}).  A spectral resolution of $R=15,000$ was chosen,
since G1's large velocity dispersion renders a higher resolution
unnecessary.  The 600 gr/mm cross disperser was used at a central
wavelength of 5822 \AA; as a result the wavelength coverage is $\sim
4800-5790$~\AA \hspace{0.025in} and $\sim 5830-6820$~\AA.
Simultaneous sky spectra were obtained with adjacent sky fibres
located 10\arcsec$\;$ from the cluster centre.  G1's half-light radius
($r_h = 1.73$\arcsec; \citealt{Ma2007}) is less than the size of the
HRS fibre (3\arcsec).  \cms{However, note that the sky fibres are
contaminated by star light, at least one of them by a bright,
foreground star.}  For that reason, sky spectra were not
subtracted from the target spectrum. \cms{Since G1 is bright,
  individual exposures were short (15 minutes), the spectra do not
  extend far into the blue, and the remaining sky fiber had minimal
  flux, the sky continuum will not have a significant effect on the
  final IL spectrum or the subsequent analysis.}

\begin{table}
\centering
\caption{Information about G1.\label{table:Targets}}
  \begin{tabular}{@{}lcc@{}}
  \hline
Parameter & Value & Note/Reference \\
  \hline
RA (J2000)  & 00:32:46.536 & 1 \\
Dec (J2000) & $+$39:34:40.67 & 1 \\
$V_{\rm{tot}}$ & 13.81 & 2\\
 & & \\
Observation Dates & 2007 Aug 8,  & HET Spectrum \\
                  & 2008 Oct 29  & \\
Total exposure time     & 5400 s & \\
S/N (5000 \AA)$^{a}$ & 230 & \\
S/N (6500 \AA)$^{a}$ & 320 & \\
$v_{\rm{helio}}$ (km s$^{-1}$) & $-349.7$ & \\
$\sigma_V$ (km s$^{-1}$)      & $23.9\pm2.0$ & \\
 & & \\
Observation Dates & 2014 Oct 5               & APO Spectrum \\
Exposure time     & 900 s                    & \\
S/N (8600 \AA)$^{a}$ & 216                    & \\
$v_{\rm{helio}}$ (km s$^{-1}$) & $-335.3$ & \\
 & & \\
Literature $v_{\rm{helio}}$ (km s$^{-1}$) & $-335\pm5$ & 3 \\
Literature $\sigma_V$ (km s$^{-1}$) & $21.4\pm 1.3$ &
Uncorrected$^{b}$; 4 \\
                             & $24.5\pm 1.5$ & Corrected$^{b}; 4$ \\
\hline
\end{tabular}
\medskip
\raggedright {\bf References: }\\
1: SIMBAD; 2: \citet{RBCref}; 3: \citet{Veljanoski2014}; 4:
\citet{Cohen2006}\\
$^{a}$ S/N ratios are per resolution element.\\
$^{b}$ \citet{Cohen2006} applied a correction to the velocity
dispersion that accounted for the aperture size.\\
\end{table}


The data reduction was performed in the Image Reduction and
Analysis Facility program (IRAF).\footnote{IRAF is distributed by the
  National Optical Astronomy Observatory, which is operated by the
  Association of Universities for Research in Astronomy, Inc., under
  cooperative agreement with the National Science Foundation.}
Standard data reduction procedures for echelle spectra were adopted;
since the cluster is so bright there is no need for variance
weighting to remove cosmic rays, unlike in previous IL analyses
\citep{Sakari2013}.  The individual exposures were shifted to the rest
frame through cross-correlations with the high resolution, high S/N
Arcturus spectrum from
\citet{Hinkle2003}.\footnote{\url{ftp://ftp.noao.edu/catalogs/arcturusatlas/}}
The final, heliocentric radial velocity is shown in Table
\ref{table:Targets}.  After the spectra were shifted to the rest
frame, individual exposures were combined with average sigma-clipping
routines.  The cluster velocity dispersion (also given in Table
\ref{table:Targets}) was determined through a cross-correlation with
Arcturus, using a calibrated relationship between the full-width at
half maximum and the velocity dispersion 
(\citealt{Alpaslan2009}; \citealt{Sakari2013}).  No correction
was made to account for the aperture size (see \citealt{Cohen2006}).

The continuum was normalized carefully, since the moderate spectral
resolution and large velocity dispersion can lead to line blanketing
in regions with strong absorption.  The blaze function of 
the orders was removed with low-order polynomial fits and the
individual orders were then combined.  Because all lines are fit with
spectrum syntheses, continuum problems are not likely to be a
significant problem in smaller 10 \AA \hspace{0.025in} regions.

Figure \ref{fig:47TucComp1} shows the HRS IL spectrum of G1 compared to
the IL spectrum of 47~Tuc, where the high-resolution 47~Tuc spectrum
from \citet{McWB} has been broadened to match the velocity dispersion
of G1. The two spectra are generally quite similar, although there are
a few regions where they differ, either due to continuum issues or
differing line strengths. A more detailed comparison will be discussed
further below.

\subsection{CaT Spectrum}\label{subsec:CaTObservations}
A CaT spectrum of G1 was obtained in 2014 with the Dual Imaging
Spectrograph (DIS) on the Astrophysical Research Consortium 3.5~m
telescope at Apache Point Observatory in New Mexico.  The
observational program for these IL CaT measurements is described in
\citet{SakWall2016}.  The spectral range from $\sim~8000-9100$
\AA \hspace{0.025in} was covered by the red camera.  The R1200 grating
and 1.5$\arcsec$ slit give a spectral resolution of $R~\sim~4000$ in
the CaT region, which is sufficient to detect the strong CaT lines.
The length of the slit (6$\arcmin$) fully covered G1 past its tidal
radius (22$\arcsec$; \citealt{Ma2007}).  An exposure time of 15 min
yielded a $\rm{S/N}\sim$ 216 per resolution element at 8600~\AA.

\begin{figure}
\begin{center}
\centering
\hspace*{-0.25in}
\includegraphics[scale=0.55,trim=0in 0 0.5in 0.0in,clip]{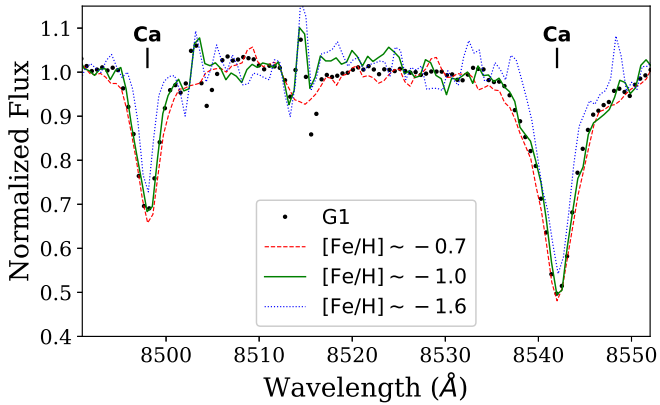}
\caption{The first two CaT lines in the G1 spectrum (black dots),
  compared to three spectra of M31 GCs from \citet{SakWall2016}: B225
  (with a CaT
  $[\rm{Fe/H}]~=~-~0.7$), B182 ($[\rm{Fe/H}]~=~-1.0$), and
  B012 ($[\rm{Fe/H}]~=~-1.6$).  G1's CaT lines are most similar to the
  B182 spectrum.}\label{fig:CaT}
\end{center}
\end{figure}

The CaT data were reduced in IRAF, as described in
\citet{SakWall2016}, utilizing variance weighting, sky subtraction
with aligned sky lines, and careful continuum normalization with a
low-order polynomical.  As with the high resolution spectrum, the
heliocentric radial velocity was determined from a cross-correlation
with the Arcturus spectrum, though in this case the Arcturus spectrum
was degraded to the resolution of DIS.  The CaT heliocentric radial
velocity agrees well with the high resolution value and the literature
value from \citet{Cohen2006}.  A portion of the CaT spectrum is shown
in Figure \ref{fig:CaT}, along with three other M31 GCs (from
\citealt{SakWall2016}).

\section{A Traditional Integrated Light Analysis: One Age, One [Fe/H]}\label{sec:Isochrone}
The first step of a traditional high-resolution IL spectral analysis,
according to the methods of \citet{McWB}, is to create a
Hertzsprung-Russell Diagram (HRD) for the cluster's underlying stellar
population.  Without high quality, resolved photometry that covers the
full cluster past the main sequence turnoff, the population must be
modeled with an isochrone, for which the age and metallicity are the
primary parameters.  A complex cluster like G1 could have stars with a
range of ages and metallicities; however, the simplest initial step is
to identify a single age and metallicity to represent the entire G1
population. The resulting effects of abundance spreads, including the
effects on the adopted models, will be investigated in a subsequent
paper.  An initial estimate of the metallicity is first obtained from
the medium-resolution CaT spectra (Section \ref{subsec:CaTFeH}); the
final parameters are then determined from a detailed inspection of the
\ion{Fe}{1} lines (Section \ref{subsec:FeH}).  Abundances and
systematic errors are then presented in Sections \ref{subsec:Abunds}
and \ref{subsec:SysErrors}.

\subsection{Age and Metallicity}\label{subsec:AgeMet}
\subsubsection{[Fe/H] from the calcium-II triplet}\label{subsec:CaTFeH}
IL CaT features have been used for decades to infer GC metallicities
(e.g., \citealt{AZ88,Foster2010,Foster2011,Usher2012}).
\citet{SakWall2016} present a comparison of CaT strengths with
high-resolution IL [Fe/H] in M31 GCs, verifying that the IL CaT is a
tracer of cluster metallicity, at least for GCs older than $\sim 3$
Gyr and with $[\rm{Fe/H}]~\la~-0.2$.\footnote{\citet{SakWall2016}
  focused on trends with [Fe/H], noting that a cluster's [Ca/Fe] can
  also affect  the strength of the CaT lines.  \citet{Usher2019} find
  that the CaT lines may track [Ca/H] better than [Fe/H].}  The precise
relationship between CaT strength and [Fe/H] depends strongly on how
the lines are measured and how the continuum is treated.  In this
paper, G1's CaT spectrum is analyzed in the same way as the other M31
GCs in \citet{SakWall2016}: its observed spectrum is fit with a linear
combination of template spectra (from stars that were observed with
the same instrumental setup) utilizing the penalized pixel-fitting
code pPXF \citep{pPXFref}. Voigt profiles were then fit to each of the
three lines to determine EWs, using the {\tt pymodelfit}
program.\footnote{\url{https://pythonhosted.org/PyModelFit/}} The
relationships between CaT EW and [Fe/H] from Table 2 in
\citet{SakWall2016} were then utilized to determine G1's CaT
metallicity, using all three CaT lines.  The CaT-based metallicity is
found to be $[\rm{Fe/H}]~=~-0.85~\pm~0.10$.  Figure \ref{fig:CaT}
compares G1's CaT spectrum with three other M31 GCs (from
\citealt{SakWall2016}), showing that G1 does indeed appear to be
moderately metal-poor.


\subsubsection{Age and Metallicity from \ion{Fe}{1} Lines}\label{subsec:FeH}
For the high-resolution analysis, the Bag of Stellar Tracks and
Isochrones (BaSTI) models from the Teramo group
\citep{BaSTIREF,BaSTIREF2} were used to synthesize G1's underlying
population, assuming an extended asymptotic giant branch (AGB) with a
mass loss parameter of $\eta = 0.2$ (see \citealt{Sakari2014} for
discussions of the effects of AGB morphology on IL analyses). The
isochrones were populated using a \citet{Kroupa2002} initial mass
function (IMF) and the resulting HRDs were binned so that each box
contains 3\% of the total flux. Alpha-enhanced (AODFNEW) Kurucz model
atmospheres\footnote{\url{http://kurucz.harvard.edu/grids.html}}
\citep{KuruczModelAtmRef} were assigned to each box, using the
$T_{\rm{eff}}$ and $\log g$ interpolation scheme from
\citet{McWB}. Following the procedure from \citet{McWB},
\citet{Colucci2009,Colucci2011,Colucci2014,Colucci2017}, and
\citet{Sakari2013,Sakari2015,Sakari2016}, an appropriate isochrone age
and metallicity were selected based on the abundances from the
\ion{Fe}{1} lines.  As explained in \citet{McWB}, this technique is
based on the method used to derive atmospheric parameters for
individual stars, and requires a sample of \ion{Fe}{1} lines that span
a range of wavelengths, reduced equivalent widths
(REW),\footnote{REW$=\log(\rm{EW}/\lambda$).} and excitation
potentials (EPs).  Since nearly every line is a blend in G1's
spectrum, equivalent widths cannot be measured for the \ion{Fe}{1}
lines; instead, each \ion{Fe}{1} line was synthesized with 20
different isochrones spanning ages of 6, 8, 10, 12, and 14 Gyr and
metallicities of $[\rm{Fe/H}]=-0.6$, $-0.7$, $-1.01$, and $-1.31$
dex. The {\it synpop} routine in the 2015 version of the Local
Thermodynamic Equilibrium (LTE) line analysis code {\tt MOOG}
\citep{Sneden} was used to synthesize spectral lines. The linelists
were generated with the {\tt linemake}
code,\footnote{\vspace{0.5in}\url{https://github.com/vmplacco/linemake}}
including hyperfine structure (HFS) and isotopic splitting as well as
molecular lines from CH, C$_{2}$, and CN.

Many of the spectral lines that can be measured in the G1 spectrum are
necessarily fairly strong, due to the spectral resolution. In general,
strong lines are undesirable for model atmospheres analysis because of
difficulties modeling the outer layers of the model atmospheres
\citep{McWilliam1995}. Unfortunately, the desirable weak lines are not
easily detectable in the broadened G1 spectrum. Line strengths were
therefore restricted to REW$<-4.6$.  This limit is higher than the
REW$=-4.7$ limit recommended by \citet{McWilliam1995}; however,
comparisons with the same lines in 47~Tuc (Section
\ref{subsubsec:LitComp}) demonstrate that selecting these higher REW
lines does not lead to significant systematic effects in the
abundances.  It is also worth remembering that although a REW limit is
placed on the IL spectral lines, the individual HRD bins could have
lines that are stronger than this REW limit; the usage of a REW limit
is therefore not straight-forward in IL analyses.

\begin{figure*}
\begin{center}
\centering
\includegraphics[scale=0.6,trim=1.2in 0 1.1in 0.2in,clip]{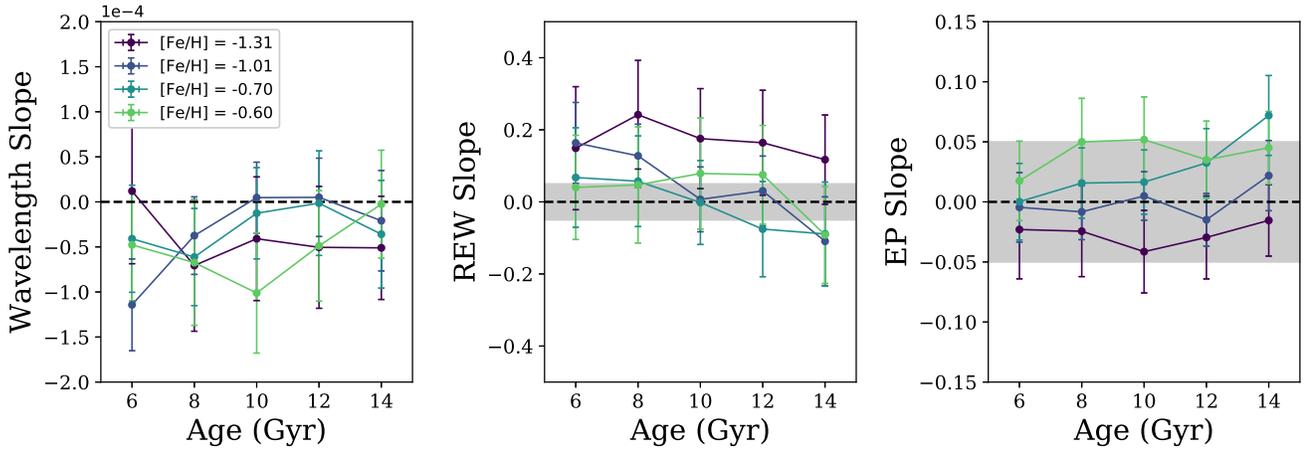}
\caption{The overall slopes of the distributions in wavelength, REW,
  and EP versus \ion{Fe}{1} abundance as a function of isochrone age
  (in Gyr). Four different isochrone metallicities are shown:
  $[\rm{Fe/H}]=-1.31$ (purple), $[\rm{Fe/H}]=-1.01$ (dark blue),
  $[\rm{Fe/H}]=-0.70$ (light blue), and $[\rm{Fe/H}]=-0.60$ (green).
  The grey bands show slopes of $\pm0.05$ in REW and
  EP, which are considered to be acceptably flat slopes.  Some of the
  variations with age amongst isochrones of the same metallicity occur
  as lines move in or out of the acceptable line strength limit of
  $\rm{REW}<-4.6$.}\label{fig:AllTrends}
\end{center}
\end{figure*}

\begin{figure}
\begin{center}
\centering
\hspace*{-0.25in}
\includegraphics[scale=0.6,trim=0.2in 0 0.6in 0.4in,clip]{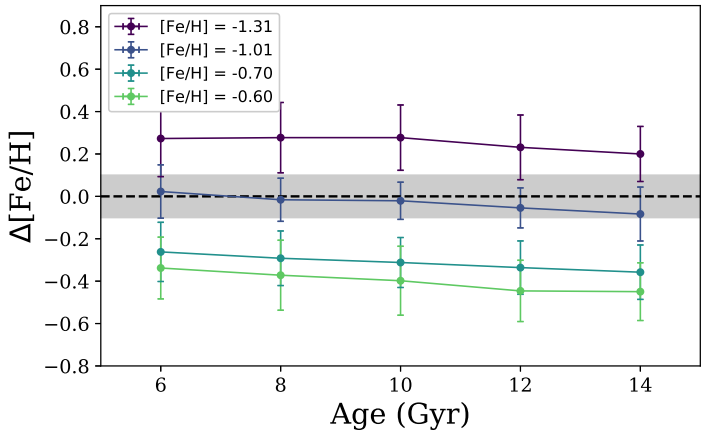}
\caption{Offsets between the average [\ion{Fe}{1}/H] and the input
  isochrone [Fe/H], as a function of isochrone age (in Gyr).  Lines
  are as in Figure \ref{fig:AllTrends}.  The grey bar shows a
  difference of $\pm0.1$ dex.}\label{fig:FeHOffset}
\end{center}
\end{figure}

\begin{figure*}
\begin{center}
\centering
\hspace*{-0.4in}
\includegraphics[scale=0.5,trim=1.6in 0 1.5in 0.5in,clip]{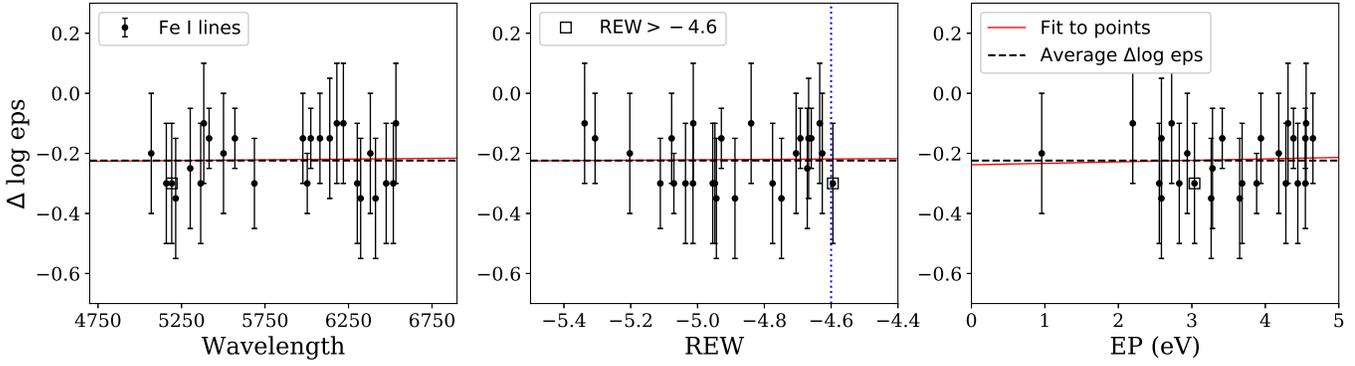}
\caption{The $\Delta[\rm{Fe/H}]$ offsets as a function of wavelength,
  REW, and EP between G1 and 47~Tuc for individual \ion{Fe}{1} lines
  (circles) when the $[\rm{Fe/H}]=-1.01$, 10 Gyr isochrone is adopted
  for G1.  The vertical dotted blue line shows $\rm{REW}=-4.6$
  limit; the square shows one spectral line that falls above this
  limit and is therefore not included in the fits.  The dashed black
  lines show the average $\Delta [\rm{Fe/H}]$, while the solid red
  lines show the fits for each panel.}\label{fig:Trends}
\end{center}
\end{figure*}

Given G1's similarity to 47~Tuc, the broadened 47~Tuc spectrum was
re-analyzed, using the isochrone parameters identified by \citet[see
  Table \ref{table:Params}]{SakariThesis}.  The only lines that were
utilized for this analysis are lines that are also measured in the G1
spectrum.  The G1 Fe abundances were then considered differentially
with respect to the 47~Tuc lines.  Given that strong lines are used
both for 47~Tuc and G1, this differential analysis should reduce the
effects from uncertain atomic data, damping constants, line blends,
etc. The final trends in [\ion{Fe}{1}/H], relative to 47~Tuc, with
respect to wavelength, REW, and EP were then calculated for the twenty
isochrones, over a range of ages and metallicities.  The resulting
slopes are shown in Figure \ref{fig:AllTrends}.  The offset in [Fe/H]
between the input isochrone and the average of the {\tt MOOG} output
for the synthesized lines is shown in Figure \ref{fig:FeHOffset}.
Although a number of isochrones produce reasonably flat trends in
[Fe/H] with wavelength, REW, and EP, the best match is for an
isochrone with $[\rm{Fe/H}]=-1.01$ dex and an age of 10 Gyr (see
Figure \ref{fig:Trends}).  Figures \ref{fig:AllTrends} and
\ref{fig:FeHOffset} also show that the cluster age is poorly
constrained, which is consistent with previous results from
\cite{Colucci2009,Colucci2014}, \cite{Sakari2015}, and other papers,
who found that the optical \ion{Fe}{1} lines were not very sensitive
to the cluster age.

\begin{table}
\centering
\begin{center}
\caption{Parameters of the adopted single-population isochrones, as
  determined from the IL spectra.\label{table:Params}}
  \begin{tabular}{@{}lccc@{}}
  \hline
        & \multicolumn{3}{c}{Isochrone Parameters} \\
Cluster & [Fe/H] & Age (Gyr) & [$\alpha$/Fe]\\
  \hline
G1     & -1.01 & 10 & $+0.4$\\
47~Tuc & -0.70 & 10 & $+0.4$\\
\hline
\end{tabular}
\end{center}
\medskip
\raggedright .\\
\end{table}

\begin{figure*}
\begin{center}
\centering
\hspace*{-0.25in}
\includegraphics[scale=0.6,trim=1.2in 0 1.0in 0.0in,clip]{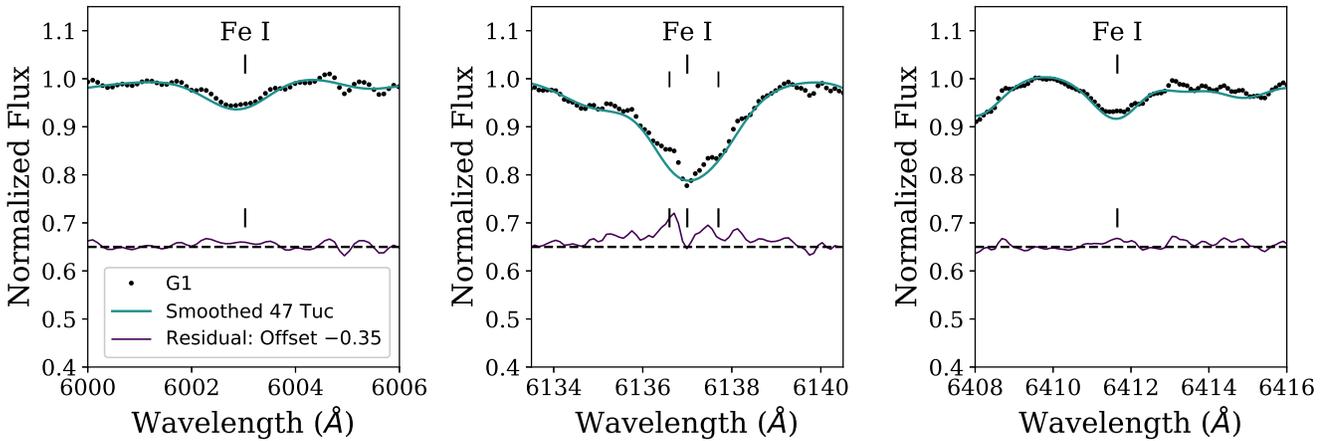}
\caption{A comparison between three \ion{Fe}{1} features in the
  optical, high-resolution IL spectra of G1 (points) and 47 Tuc (bold
  green line; from \citealt{McWB}), where the 47~Tuc spectrum has been
  smoothed to the velocity dispersion of G1.  The small vertical lines
  show the locations of individual spectral lines. The thin purple
  line (arbitrarily offset below the two spectra) shows the
  residuals. These small offsets indicate that G1 is slightly more
  metal-poor than 47~Tuc.}\label{fig:47TucFeComp}
\end{center}
\end{figure*}

The average metallicity for G1 is slightly more metal-poor than 47~Tuc
($[\rm{Fe\;I/H}]~=~-0.98~\pm~0.05$ dex compared to
$[\rm{Fe\;I/H}]~=~-0.76~\pm~0.03$ dex).  Figure \ref{fig:47TucFeComp}
shows that several of the lines in G1 are indeed weaker than the lines
in 47~Tuc, supporting a lower metallicity. The high-resolution
spectroscopic metallicity is also lower than the IL CaT-based [Fe/H]
(Section \ref{subsec:CaTFeH}).  This discrepancy may reflect
$\alpha$-enhancement in G1.  With a large sample of GCs in the Milky
Way and nearby dwarf galaxies, \citet{Usher2019} find that the CaT
strength is a better tracer of [Ca/H] than [Fe/H].  For a cluster with
$[\rm{Ca/Fe}] = +0.4$, they find that an [Fe/H] derived from the CaT
could be about $0.1$ dex higher than the value from a high-resolution
analysis.  If G1 is similarly $\alpha$-enhanced (see Section
\ref{subsubsec:Alpha}), then we might expect that its CaT [Fe/H] has
been over-estimated by $0.1$ dex.

\subsubsection{G1's Metallicity: Comparisons with Previous Results}\label{subsec:FeHComp}

The spectroscopic metallicity of $[\rm{Fe/H}] = -0.98\pm0.05$ is
consistent with previous photometric [Fe/H] estimates for G1.  Based
on measurements of the RGB slope in the {\it HST} photometry and
comparisons with 47~Tuc fiducials, \citet{Rich1996} and
\citet{Meylan2001} find that G1 should have an average [Fe/H] around
$-0.95$.  Based on a similar analysis, \citet{Federici2012} find
$[\rm{Fe/H}]~=~-0.90$. \citet{Meylan2001} further concluded that there
could be a wide Fe spread of $0.4-0.5$ dex based on the width of G1's
RGB. \citet{Nardiello2019} conducted a subsequent re-analysis of the
{\it HST} photometry, focusing on signs of abundance spreads within G1.
Based on comparisons with the BaSTI isochrones they adopted a standard
red RGB sequence with $[\rm{Fe/H}] = -0.70$ and found that a much
smaller spread (down to $[\rm{Fe/H}] = -0.85$) was consistent with the
observed width of the RGB---they argue that the $\Delta[\rm{Fe/H}]$
could be slightly smaller if there are additional helium or CNO
variations within the cluster. G1's primarily red horizontal branch
also indicates that the dominant population is fairly
metal-rich---however, many authors
\citep{Rich1996,Meylan2001,Perina2012,Nardiello2019} have noted the
presence of bluer horizontal branch stars, which further indicates
that there may be a helium spread within G1.

The spectroscopic [Fe/H] derived here also agrees with previous
spectroscopic analyses.  \citet{Reitzel2004} obtained CaT spectra of
individual M31 field stars, including at least one star that is a
likely member of G1 based on its radial velocity; this likely member
has a CaT-based metallicity of $[\rm{Fe/H}]~=~-0.74$.  Other IL
spectroscopic analyses at lower-resolution also find similar
metallicities.  The earliest abundance analysis by
\citet{vandenBergh1969} found $[\rm{Fe/H}]~=~-0.8$, while subsequent
analyses found similar results (e.g., \citealt{Huchra1991} found
$[\rm{Fe/H}]~=~-1.01$).  The Revised Bologna Catalog
(RBC)\footnote{\url{http://www.bo.astro.it/M31/}} reports
$[\rm{Fe/H}]~=~-0.73\pm0.15$, based on a Lick index analysis
\citep{RBCref}---however, \citet{Colucci2014} note that the Lick
index [Fe/H] ratios of their sample of M31 GCs are slightly higher
than the high-resolution values for clusters at $[\rm{Fe/H}]~=~-1$.
The spectroscopic value in this paper is therefore generally
consistent with other values from the literature.

Of course, the possibility of an iron spread in G1 makes it difficult
to interpret a single IL value.  The metallicity of G1 will be
discussed more in Section \ref{subsubsec:Fe} and in a forthcoming
paper.

\subsection{Detailed Abundances of G1}\label{subsec:Abunds}
The abundances of other elements were determined via spectrum
syntheses in {\tt MOOG}, using the single stellar population isochrone
parameters in Table \ref{table:Params}.  Note that for IL analyses
\citet{Sakari2013} found that it was better to use line-to-line
differential abundances with respect to solar abundances derived with
the same techniques, atomic data, etc.  That technique has not been
used here, however, since many of the lines that are detectable in the
G1 spectrum are too strong in the solar spectrum.  Instead, 47~Tuc is
used as a reference.  Comparisons with literature values for 47~Tuc
are given in Section \ref{subsubsec:LitComp}.

The single population [Fe/H] and [X/Fe] abundance ratios are shown in
Table \ref{table:Abunds}; individual abundances per spectral line are
given in Appendix \ref{appendix:Abunds}.  Abundances for 47~Tuc are
also shown in Table \ref{table:Abunds}, along with the abundance ratio
differences between G1 and 47~Tuc.

\begin{table*}
\centering
\begin{center}
\caption{Isochrone-based IL Abundances for G1 and 47~Tuc.\label{table:Abunds}}
  \newcolumntype{d}[1]{D{,}{\;\pm\;}{#1}}
  \begin{tabular}{@{}ld{6}ccd{6}cd{6}@{}}
  \hline
        & \multicolumn{2}{l}{G1} & & \multicolumn{2}{l}{47~Tuc} &
  \multicolumn{1}{l}{$\Delta$Abundance}\\
Element & \multicolumn{1}{l}{Abundance} & $N$ & &
\multicolumn{1}{l}{Abundance} & $N$ & \multicolumn{1}{l}{(G1$-$47 Tuc)} \\
  \hline
$[$\ion{Fe}{1}/H$]$ & -0.98,0.05 & 25 & & -0.76,0.03 & 25  & -0.22 \\
$[$\ion{Fe}{2}/H$]$ & -0.83,0.10 & 2  & & -0.70,0.10 & 2   & -0.13 \\
$[$\ion{C}{1}/Fe$]$ & \multicolumn{1}{c}{$<0.17$} & 4 & & \multicolumn{1}{c}{$<-0.44$} & 4 & \multicolumn{1}{c}{--$\;\;\;\;\;\;\;\;\;$}\\
$[$\ion{Na}{1}/\ion{Fe}{1}$]$ &  0.60,0.13 & 4 & & 0.38,0.05 & 4  & +0.22 \\ 
$[$\ion{Mg}{1}/\ion{Fe}{1}$]$ &  0.38,0.12 & 2 & & 0.41,0.11 & 2  & -0.03 \\
$[$\ion{Al}{1}/\ion{Fe}{1}$]$ &  0.72,0.20 & 2 & & 0.31,0.08 & 2  & +0.41 \\
$[$\ion{Ca}{1}/\ion{Fe}{1}$]$ &  0.36,0.07 & 9 & & 0.32,0.04 & 8  & +0.04 \\
$[$\ion{Ti}{1}/\ion{Fe}{1}$]$ &  0.34,0.08 & 4 & & 0.29,0.07 & 4  & +0.05 \\
$[$\ion{Ti}{2}/\ion{Fe}{1}$]$ &  0.27,0.09 & 2 & & 0.34,0.08 & 2  & -0.07 \\
$[$\ion{Ti}{2}/\ion{Fe}{2}$]$ &  0.12,0.15 & 2 & & 0.28,0.13 & 2  & -0.16 \\
$[$\ion{Cr}{1}/\ion{Fe}{1}$]$ & -0.13,0.06 & 2 & & -0.14,0.07 & 2 & +0.01 \\
$[$\ion{Mn}{1}/\ion{Fe}{1}$]$ & -0.16,0.12 & 3 & & -0.22,0.14 & 3 & +0.06 \\
$[$\ion{Ni}{1}/\ion{Fe}{1}$]$ &  0.02,0.09 & 2 & & 0.01,0.08 & 2  & +0.01 \\
$[$\ion{Cu}{1}/\ion{Fe}{1}$]$ & -0.63,0.20 & 1 & & -0.34,0.10 & 1 & -0.29 \\
$[$\ion{Zn}{1}/\ion{Fe}{1}$]$ &  0.27,0.15 & 1 & & 0.06,0.10 & 1  & +0.21 \\
$[$\ion{Y}{2}/\ion{Fe}{1}$]$  & -0.13,0.15 & 2 & & -0.04,0.09 & 2 & -0.09 \\
$[$\ion{Y}{2}/\ion{Fe}{2}$]$  & -0.28,0.17 & 2 & & -0.10,0.13 & 2 & -0.18 \\
$[$\ion{Ba}{2}/\ion{Fe}{1}$]$ &  0.00,0.11 & 3 & & 0.16,0.08 & 3  & -0.16 \\  
$[$\ion{Ba}{2}/\ion{Fe}{2}$]$ & -0.15,0.16 & 3 & & 0.10,0.13 & 3  & -0.25 \\
$[$\ion{Eu}{2}/\ion{Fe}{1}$]$ & \multicolumn{1}{c}{$<0.49$} & 1 & &  0.30,0.15 & 1  & \multicolumn{1}{c}{--$\;\;\;\;\;\;\;\;\;$} \\
$[$\ion{Eu}{2}/\ion{Fe}{2}$]$ & \multicolumn{1}{c}{$<0.32$} & 1 & &  0.36,0.18 & 1  & \multicolumn{1}{c}{--$\;\;\;\;\;\;\;\;\;$} \\
$[$Ba/Y$]$                    &  0.13,0.18 & --  & & 0.20,0.12 & -- & -0.07 \\
$[$Ba/Eu$]$                   & \multicolumn{1}{c}{$>-0.49$} & -- & &
  -0.14,0.17 & --  & \multicolumn{1}{c}{--$\;\;\;\;\;\;\;\;\;$} \\
\hline
\end{tabular}
\end{center}
\medskip
\raggedright .\\
\end{table*}

\subsubsection{Iron}\label{subsubsec:Fe}
The 25 \ion{Fe}{1} lines used for this analysis span a range of
wavelengths, EPs, and line strengths, as discussed in Section
\ref{subsec:FeH}.  Two \ion{Fe}{2} lines, at 5534 and 6456 \AA, are
also detectable in G1's spectrum.  The resulting [\ion{Fe}{2}/H] is
0.15 dex higher than the [\ion{Fe}{1}/H] ratio.  A $+0.06$ dex offset
between \ion{Fe}{2} and \ion{Fe}{1} is also found in 47~Tuc. Other
high-resolution IL (and single star) analyses have found similar
discrepancies between \ion{Fe}{1} and \ion{Fe}{2} (e.g.,
\citealt{McWB}, \citealt{Colucci2009,Colucci2012,Colucci2014},
\citealt{Sakari2015,Sakari2016}).  Such differences may reflect the
weakness and paucity of \ion{Fe}{2} lines in IL spectra.  \citet{McWB}
also suggest that offsets between \ion{Fe}{1} and \ion{Fe}{2} could
occur as a result of incorrect modelling of the underlying stellar
population; they list two possible sources of error: 1)~a mismatch
between the [$\alpha$/Fe] ratio of the cluster and the isochrone and
2)~an incorrect number of bright asymptotic giant branch or tip of the
RGB stars in the adopted model.  Both possibilities were confirmed to
have different effects on the IL [\ion{Fe}{1}/H] and  [\ion{Fe}{2}/H]
ratios by \citet{Sakari2014}, based on tests with 47~Tuc and other
clusters.  Non-LTE (NLTE) effects can also lead to lower \ion{Fe}{1}
abundances in LTE analyses (e.g., \citealt{Lind2012,Amarsi2016});
however, these effects are not very strong at G1's metallicity. Even
if the underlying stellar population in G1 has not been perfectly
modelled, the [X/Fe] ratios are less sensitive to these effects than
[X/H] ratios (\citealt{Sakari2014}, though see Section
\ref{subsec:SysErrors}).

\subsubsection{Carbon}\label{subsubsec:CFe}
The C abundance is difficult to ascertain from this spectrum, since
the molecular lines are relatively weak and blended at this spectral
resolution.  Upper limits on [C/Fe] were determined from C$_2$
features at 5135, 5165, 5585, and 5635~\myAA. The upper limit of
$[\rm{C/Fe}]~<~0.17$ suggests that the cluster is not C-enhanced.
This lower value is consistent with expected values from tip of the
RGB stars and with other IL analyses (e.g., \citealt{Schiavon2013};
\citealt{Sakari2016}).


\subsubsection{Sodium and Aluminum}\label{subsubsec:Na}
\begin{figure}
\begin{center}
\centering
\hspace*{-0.1in}
\includegraphics[scale=0.55,trim=0.in 0.5in 0.2in 0.7in,clip]{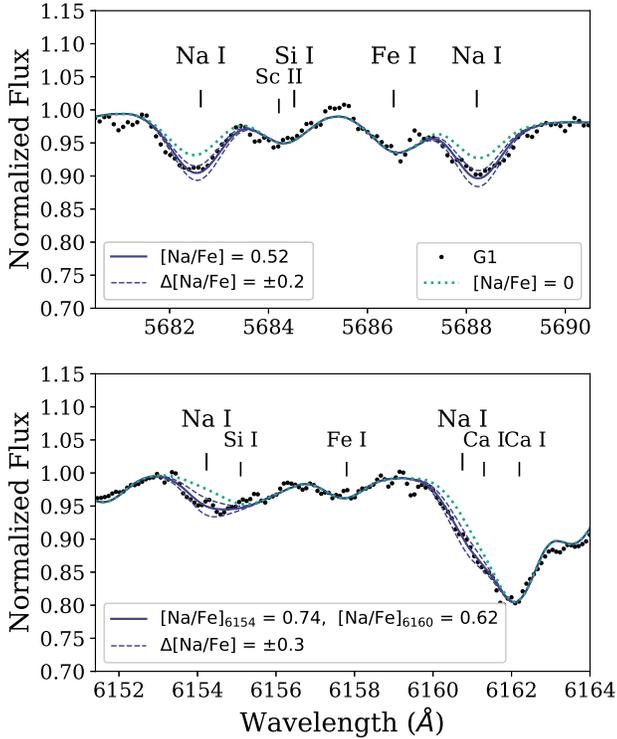}
\caption{Syntheses of the 5682/5688 \myAA (top) and 6154/6160 \myAA
  (bottom) \ion{Na}{1} doublets.  The black points show the G1
  spectrum.  The solid lines show the best-fitting syntheses, while
  the dashed lines show uncertainties of 0.2 and 0.3, respectively.
  The dotted green line shows a solar [Na/Fe] ratio; neither set of
  doublets is consistent with a solar [Na/Fe] ratio.}\label{fig:NaSynth}
\end{center}
\end{figure}

\begin{figure}
\begin{center}
\centering
\hspace*{-0.1in}
\includegraphics[scale=0.55,trim=0.in 0.5in 0.2in 0.7in,clip]{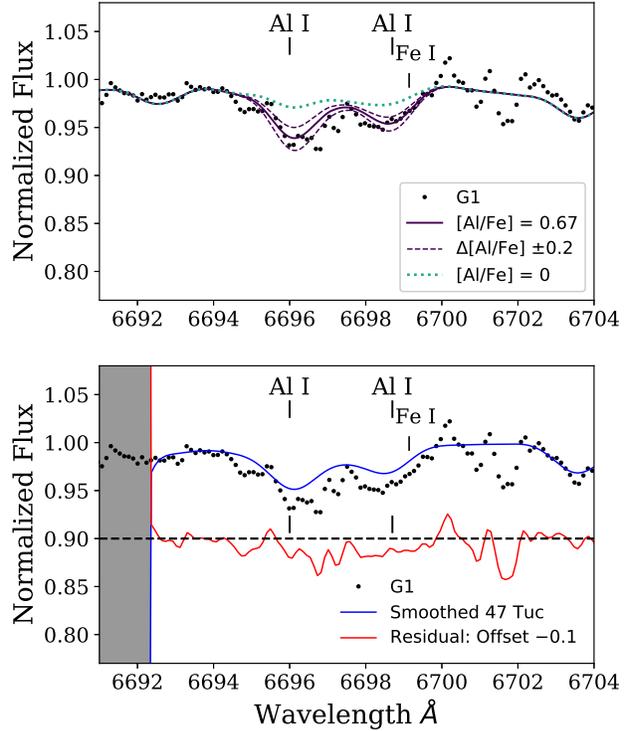}
\caption{The 6696 and 6698 \myAA \ion{Al}{1} lines in G1. The black
  points show the G1 spectrum.  {\it Top: } Spectrum syntheses of the
  Al lines.  The solid line shows the best-fitting syntheses, while
  the dashed lines show uncertainties of 0.2 and 0.3, respectively.
  The dotted green line shows a solar [Al/Fe] ratio; neither line is
  consistent with a solar [Al/Fe] ratio. {\it Bottom: } A comparison
  with the 47~Tuc spectrum (solid blue line), which has been smoothed to
  G1's velocity dispersion.  The residual is shown below the spectra,
  offset by $-0.1$ dex.  \cms{As in Figure \ref{fig:47TucComp1}, the
    shading shows a bad region of the 47~Tuc spectrum.}}\label{fig:AlSynth}
\end{center}
\end{figure}

The sodium abundances were derived from two sets of \ion{Na}{1}
doublets: the lines at 5682 and 5688 \myAA and those at 6154 and
6160~\AA.  Aluminum was derived from the 6696 and 6698 \myAA
lines. The syntheses for these features are shown in Figures
\ref{fig:NaSynth} and \ref{fig:AlSynth}; the latter figure also shows
a comparison with 47~Tuc.  Solar ratios of $[\rm{Na/Fe}] = 0$ and
$[\rm{Al/Fe}] = 0$ are also shown, demonstrating that the cluster is
enhanced in Na and Al.  Table \ref{table:Abunds} shows that these
lines lead to [Na/Fe] and [Al/Fe] ratios in G1 that are about 0.2 and
0.4 dex higher, respectively, than 47~Tuc (see Section
\ref{subsubsec:Na}).

The two sets of Na lines are known to have NLTE corrections: for a
typical RGB star in G1, the INSPECT
database\footnote{\url{http://inspect.coolstars19.com/}}
\citep{Lind2011} indicates that the corrections would be negative (up
to $\sim-0.2$ dex for the 5682/5688 \myAA lines).  Similar
corrections would also be needed in 47~Tuc.  The metallicity
difference between the two clusters could lead to different NLTE
corrections; however, the INSPECT database indicates that the
difference between NLTE corrections is not significant for
$[\rm{Fe/H}]~=~-1$ versus $-0.7$.  This indicates that G1's higher
[Na/Fe] cannot be explained solely with NLTE.  Note that these NLTE
corrections are not applied here.  \cms{The implications of these Na
  and Al abundances will be discussed further in Section
  \ref{sec:Discussion}.}

\subsubsection{The $\alpha$ elements}\label{subsubsec:Alpha}
\begin{figure}
\begin{center}
\centering
\includegraphics[scale=0.53,trim=0.1in 0 0.4in 0.3in,clip]{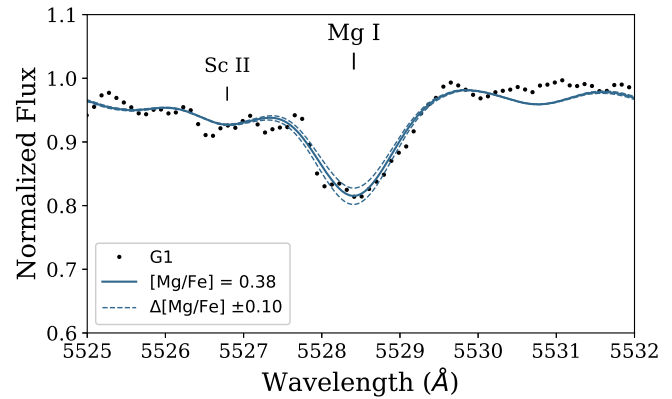}
\caption{Syntheses of the 5528 \myA (left) and 5711 \myA
  (right) \ion{Mg}{1} lines.  The black points show the G1
  spectrum.  The solid line shows the best-fitting syntheses, while
  the dashed lines show uncertainties.}\label{fig:MgSynth}
\end{center}
\end{figure}

Mg, Ca, and Ti abundances were determined from a variety of spectral
lines.  In the case of Mg, lines at 5528 and 5711~\myAA were
used---note that the 5528~\myAA line is rather strong in G1 and
47~Tuc, while the 5711~\myAA is barely detectable in G1.  Both Mg
lines in G1 yield a supersolar [Mg/Fe] ratio (see Figure
\ref{fig:MgSynth}).  Nine Ca lines were measured, spanning a broad
range in wavelength; all indicate that [Ca/Fe] is supersolar.
Finally, four \ion{Ti}{1} and two \ion{Ti}{2} lines were measured,
indicating elevated [Ti/Fe].  Note that the [\ion{Ti}{2}/\ion{Fe}{2}]
ratio is lower than the [\ion{Ti}{2}/\ion{Fe}{1}] ratio, which may
indicate a problem with the derived [\ion{Fe}{2}/H]  (see Section
\ref{subsubsec:Fe}).  Lower [\ion{Ti}{2}/\ion{Fe}{2}] ratios are not
necessarily unusual; from IL observations, \citet{Colucci2017} found
lower [\ion{Ti}{2}/Fe] ratios than [\ion{Ti}{1}/Fe] for several Milky
Way GCs. This offset between \ion{Ti}{1} and \ion{Ti}{2} may indicate
that the weaker, bluer \ion{Ti}{2} lines may be less reliable
than the \ion{Ti}{1} lines.  Ultimately, G1 appears to be
$\alpha$-enhanced, similar to 47~Tuc and the majority of the Milky Way
and M31 GCs---this will be discussed in more detail in Section
\ref{subsec:G1NSC}.

\subsubsection{Iron-peak elements, Cu, and Zn}\label{subsubsec:FePeak}
Abundances of the iron-peak elements Cr, Mn, and Ni were determined
from an assortment of spectral lines; in the case of Mn, HFS
components were also included.  Cu was determined from the 5782 \myAA
line (assuming a solar isotopic ratio; \citealt{Asplund2009}) with the
HFS components from the Kurucz
database.\footnote{\url{http://kurucz.harvard.edu/linelists.html}}
\cms{Note that the stronger 5105 \myAA \ion{Cu}{1} line was located in a
lower S/N region of the G1 spectrum, and was not included.} Zn was
determined from the 4810 \myAA line.

Cr, Mn, and Ni are considered to be standard iron-peak elements, while
Cu and Zn are though to form via weak $s$-processing
\citep{Bisterzo2004,Pignatari2010}.  Ultimately, G1's [Ni/Fe] ratio is
solar, [Cr/Fe] and [Mn/Fe] are slightly subsolar, [Cu/Fe] is
significantly subsolar, and [Zn/Fe] is slightly elevated. These trends
are generally similar to the results in 47~Tuc. Although [Cu/Fe] is
lower in G1, this is consistent with standard Milky Way chemical
evolution at $[\rm{Fe/H}]~\sim~-1$ versus $-0.8$ (see
\citealt{McWilliam2013}).  The  high [Zn/Fe] in G1 agrees with typical
Milky Way field stars within its errors \citep{Bisterzo2004}.

\subsubsection{Neutron capture elements}\label{subsubsec:NeutronCapture}
Two \ion{Y}{2} lines, at 5087 and 5200 \AA, were used to determine
[Y/Fe], while three \ion{Ba}{2} lines, at 5853, 6141, and 6496 \AA,
were used to determine [Ba/Fe].  All three of the Ba lines have
isotopic splitting; a solar isotopic ratio was assumed
\citep{Asplund2009}.  An upper limit in [Eu/Fe] for G1 was determined
from the \ion{Eu}{2} 6645 \myAA line, including isotopic components
and a Solar isotopic ratio \citep{Asplund2009}.  The resulting G1
abundances show that [Y/Fe] is slightly subsolar, [Ba/Fe] is
approximately solar (depending on whether \ion{Fe}{1} or \ion{Fe}{2}
is used), and [Eu/Fe] is moderately enhanced at most.

Y and Ba are mainly created by the slow ($s$-) neutron-capture
process, while Eu is mainly created by the rapid neutron-capture
($r$-) process \citep{Burris2000}.  The slightly elevated [Ba/Y] ratio
hints at a small excess of heavier neutron-capture elements; this
small excess is also found in 47~Tuc, and is consistent with the
general Milky Way trend.  G1's [Ba/Eu] ratio indicates
that it is consistent with being dominated by $r$-process material,
similar to 47~Tuc, though it may have received some small
contributions from the $s$-process.  The implications of the
neutron-capture abundances in G1 will be discussed further in
Section \ref{subsubsec:NeutronCaptureDiscussion}.

\subsubsection{47 Tuc: Comparison with Literature Abundances}\label{subsubsec:LitComp}
\begin{figure*}
\begin{center}
\centering
\includegraphics[scale=0.65,trim=0in 0 0.0in 0.0in,clip]{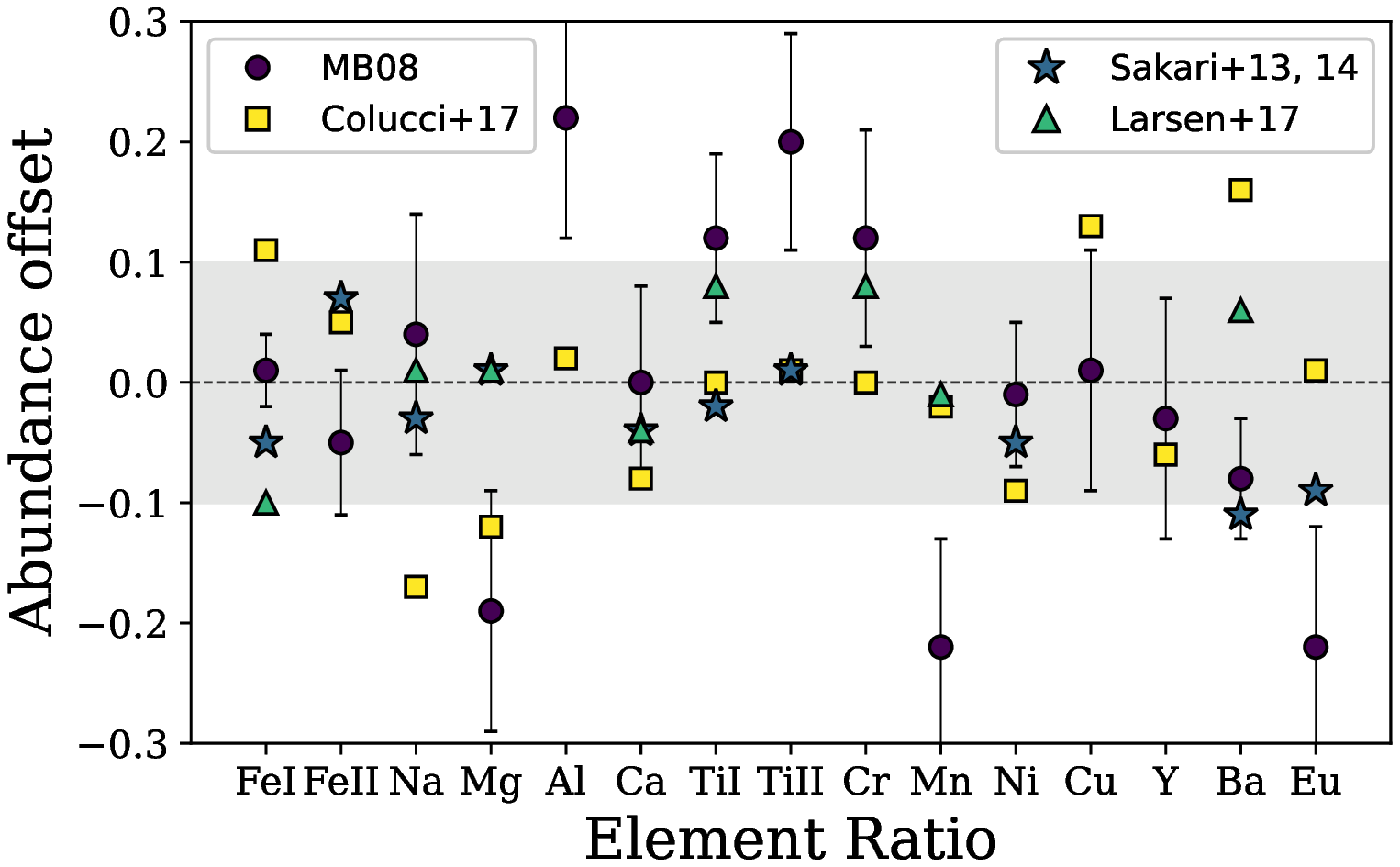}
\caption{Abundance ratio offsets \cms{for 47~Tuc} (literature $-$ this
  paper).  For \ion{Fe}{1} and \ion{Fe}{2}, [Fe/H] ratios are
  compared.  For all other elements, [X/Fe] ratios are compared;
  \ion{Fe}{2} ratios are used for [X/Fe] ratios of singly ionized
  species.  The literature data from \citet[purple circles]{McWB},
  \citet[blue stars]{Sakari2013,Sakari2014}, \citet[yellow
    squares]{Colucci2017}, and \citet[green triangles]{Larsen2017} are
  shown.  The grey bar shows a range of $\pm0.1$ dex, which is a
  typical uncertainty for the [X/Fe] ratios in this
  paper. Representative error bars are shown for the \citet{McWB}
  points only.}\label{fig:47TucComp}
\end{center}
\end{figure*}

Previous IL abundances for 47~Tuc have been derived by \citet{McWB},
\citet{Sakari2013,Sakari2014}, \citet{Colucci2017}, and
\citet{Larsen2017}.  \cms{This paper has presented a new set of IL
  abundances for 47~Tuc, using only the stronger lines that are
  detectable in G1.  Offsets between the 47~Tuc abundances in this
  analysis versus the literature can provide insight into systematic
  offsets that could be present in the G1 abundances as well.}  These
abundance offsets are shown in Figure \ref{fig:47TucComp}.

When comparing the results from this paper to these analyses from the
literature, an important caveat is that \citet{McWB},
\citet{Sakari2013,Sakari2014}, and \citet{Colucci2017} have all
analysed the same 47~Tuc spectrum that is used in this paper, but with
different techniques.  \citet{Larsen2017} obtained a separate 47~Tuc
spectrum.  \citet{McWB}, \citet{Sakari2013,Sakari2014}, and
\citet{Larsen2017} used {\it HST} photometry to model the underlying
populations---this is only possible for nearby GCs whose stars can be
resolved down to the main sequence. \citet{Colucci2017} used
theoretical isochrones to model the underlying populations, similar to
the analysis here.  For the abundance analysis, \citet{McWB}
performed an equivalent width (EW) analysis,
\citet{Sakari2013,Sakari2014} and \citet{Colucci2017} used a mixture
of EWs and spectrum syntheses, and \citet{Larsen2017} used
full-spectrum fitting with synthetic spectra.  The choice of spectral
lines and atomic data also vary between these papers.  Although the
analysis in this paper uses the same spectrum as \citet{McWB} and the
subsequent analyses, this paper only uses the spectral lines that are
detectable in G1.

Figure \ref{fig:47TucComp} shows that most of the points fall within
the grey bar, which indicates the typical uncertainty for the [X/Fe]
ratios in this paper.  There are likely to be systematic errors
between the analyses as a result of the differences in modeling the
underlying stellar populations (see, e.g., \citealt{Sakari2014}).
One major difference is that, because of the requirement to only use
lines that are detectable in G1, this analysis uses stronger lines
than \citet{McWB}, \citet{Sakari2013,Sakari2014}, or
\citet{Colucci2017}.  A few of the key outliers Figure
\ref{fig:47TucComp} are discussed below.  The high-precision,
differential abundances of individual 47~Tuc stars from
\citet{KochMcW} are also included in this discussion.

\begin{description}
\item[Fe I: ] Although the [\ion{Fe}{1}/H] ratio derived in this paper
  is generally in agreement with the literature analyses,
  \citet{Colucci2017} find a higher [Fe/H], by $0.11$~dex
  ($[\rm{Fe/H}]~=~-0.65$).  This may be due to the line selection,
  since stronger lines are utilized in this analysis; however, 12 of
  the 25 lines in this analysis are also included in
  \citet{Colucci2017}.  With a similar line selection as Colucci et
  al., \citet{SakariThesis} found $[$\ion{Fe}{1}/H$]~=~-0.74\pm0.03$
  with theoretical isochrones, which is in better agreement with the
  [Fe/H] derived here.
  
  In a differential analysis of nine individual stars in
  47~Tuc, \citet{KochMcW} found a metallicity of
  $[$\rm{\ion{Fe}{1}/H}$]~=~-0.76$, with a random error of 0.01 dex and a
  systematic error of 0.04 dex. The result from this paper is in
  excellent agreement with this high-quality values from individual
  stars.
\item[Na: ] \citet{Colucci2017} find a Na abundance of
  $[\rm{Na/Fe}]~=~0.24~\pm~0.08$, which is $0.14$ dex lower than the
  value in this paper.  The $\log \epsilon$ abundances from Colucci et
  al. are very similar to the values in this paper; instead, this
  offset is likely due to the way the [Na/Fe] is calculated.
  \citet{Colucci2017} perform a differential analysis, relative to the
  solar abundance derived from the same lines.  A differential
  analysis is not performed here, since the 5682 and 5688 \myAA lines
  are prohibitively strong in the solar spectrum \citep{Sakari2013}.
  With the \citet{Asplund2009} solar value for Na, Colucci et al.'s
  [Na/Fe] ratio is in closer agreement with the value derived here.

  From individual stars, \citet{KochMcW} found a mean
  $[\rm{Na/Fe}]~=~0.22~\pm~0.03$ dex.  Although the IL [Na/Fe] is
  higher than this value from individual stars, this is likely due to
  the presence of Na-enhanced stars within the cluster (see
  discussions in \citealt{Sakari2013}, \citealt{Colucci2017}, and
  Section \ref{subsubsec:Na}).
\item[Mg: ] \citet{McWB} and \citet{Colucci2017} both find a lower
  [Mg/Fe] ratio for 47~Tuc.  \citet{Sakari2013} demonstrated that this
  may be due to an issue with the atomic data for the 7387 \myAA line,
  which is the only \ion{Mg}{1} line used by \citet{McWB} and one of
  three lines used by \citet{Colucci2017}.  The value derived in this
  paper is in excellent agreement with the mean value for the
  individual stars analyzed by \citet{KochMcW},
  $[\rm{Mg/Fe}]~=~+0.46\pm0.05$.
\item[Al: ]  The abundances derived here are in agreement with
  \citet{Colucci2017}, who used spectrum syntheses.  However, with EW
  analyses \citet{McWB} derive a higher [Al/Fe] of $+0.53$ for
  47~Tuc.  \cms{The discrepancy with \citet{McWB} is likely due to
    differences in the atomic data used for the 6696 and 6698 \myAA
    lines.  Adopting the $\log gf$ values from \citet{McWB} would
    increase the 47~Tuc (and G1) [Al/Fe] ratios.}  From individual
  stars, \citet{KochMcW} find an average
  $[\rm{Al/Fe}]~=~+0.45\pm0.06$, which is higher than the value
  derived in this paper.
\item[Ti: ] \citet{McWB} find higher Ti I and Ti II ratios than those
  in this analysis.  However, there is only one line in common for
  each ratio. This offset may therefore reflect uncertainties in
  atomic data or their EW measurements.
\item[Cr and Mn: ] \citet{McWB} also find higher [Cr/Fe] and lower
  [Mn/Fe] ratios, compared to this analysis. Again, these offsets may
  be due to issues with measuring EWs, or they could reflect
  systematic differences in spectral lines (no Cr lines are similar,
  while only one Mn line is in common). \cms{In particular,
    differences in NLTE corrections between the various lines could
    lead to offsets in the final [Cr/Fe] and [Mn/Fe] ratios.}
\item[Cu: ] \citet{Colucci2017} find a higher, solar [Cu/Fe] ratio for
  47~Tuc. This discrepancy seems to caused by their inclusion of the
  5105 \myAA line.  When only the 5782 \myAA is considered, the ratios
  are in agreement.
\item[Ba: ] \citet{McWB} and \citet{Sakari2014} found lower [Ba/Fe]
  ratios than this analysis, which may be a result of using EWs rather
  than syntheses. On the other hand, \citet{Colucci2017} find a higher
  [Ba/Fe] ratio than the value in this paper, possibly as a result of
  different adopted isotopic ratios. 
\item[Eu: ] \citet{McWB} also find a lower [Eu/Fe] ratio than the
  analysis in this paper; however, this may be because they used an EW
  rather than spectrum synthesis to derive the abundance (see the
  discussion in \citealt{Sakari2013}).
\end{description}

Most of the offsets in Figure \ref{fig:47TucComp} can therefore be
explained with differences in line selection (leading to offsets
because of, e.g., uncertain atomic data or solar ratios) or line
measurement techniques (EWs versus spectrum syntheses).  This
comparison between the 47~Tuc analyses demonstrates that, in general,
the usage of stronger spectral lines has not led to obvious systematic
offsets in the [X/Fe] ratios.  This result is encouraging for the IL
abundances of G1, whose detectable lines are similarly strong.
\cms{The offsets in comparison with the literature abundances will be
  also present in the abundances of G1.}

\subsection{Systematic Errors}\label{subsec:SysErrors}
There are many systematic effects that can alter the derived IL
abundance ratios.  A full systematic error exploration is beyond the
scope of this paper; a more detailed discussion of, e.g., the effects
of AGB models, HB morphology, and isochrone binning is presented in
\citet{Sakari2014}.  However, given the comparisons with 47~Tuc that
are present throughout this paper, it is worth briefly discussing
three specific sources of uncertainty in the isochrone modeling that
were identified by \citet{McWB} as being particularly problematic for
47~Tuc: M giants, mass segregation, and AGB bump stars.  Cool M giants
have significant TiO absorption that can affect IL spectral features,
particularly at red wavelengths. 47~Tuc is known to have two M giants
in its core; however, \citet{McWB} found that these two M giants have
a negligible effect on the IL spectrum, causing only 0.02 and 0.01 dex
offsets in [\ion{Fe}{1}/H] and [\ion{Fe}{2}/H], respectively.  M
giants become increasingly more prevalent in metal-rich clusters;
since G1 is more metal-poor than 47~Tuc, it is unlikely to have a
significant population of M giants (though a forthcoming paper will
discuss how an iron spread could lead to M giants in G1).  The core
region of 47~Tuc (which was covered in the IL spectrum used in this
paper) is also known to suffer from mass segregation, which removes
the lowest mass cluster stars.  \citet{Sakari2014} found that adopting
a low-mass cutoff did not have a significant effect on most abundance
ratios. Finally, \citet{McWB} found that the BaSTI isochrones
underpredicted the number of AGB bump stars in the 47~Tuc core.  An
incorrect number of AGB bump stars can have a significant effect on
the derived abundances, since AGB stars are so bright. \citet{McWB}
found that not including the extra AGB stars led to higher
[\ion{Fe}{1}/H] and [\ion{Fe}{2}/H] ratios by $\sim 0.1$ and $0.15$
dex, respectively. The exact abundance offsets will depend on the
precise modeling of the AGB, the adopted cluster age, the spectral
lines used, etc.  The exact numbers of AGB bump stars may also vary
from cluster to cluster.  Such factor are difficult to model for G1.

In addition to these specific effects, there are two sources of
systematic errors that are especially pertinent to G1: 1) the
systematic errors that occur due to uncertainties in identifying an
appropriate isochrone (this section) and 2) offsets that occur as a
result of undetected abundance spreads within the cluster.  The second
source of uncertainty will be addressed in a forthcoming paper (Sakari
et al., {\it in prep.}), while the first type is quantified
below. Table \ref{table:SysErrors} shows the abundance ratio
uncertainties that occur when the isochrone age is changed by
$\pm4$~Gyr, the metallicity by $\pm0.3$ dex, and the [$\alpha$/Fe]
ratio by $-0.4$. For neutral species all [X/Fe] ratios utilize
\ion{Fe}{1}; for singly ionized species both [X/\ion{Fe}{1}] and
    [X/\ion{Fe}{2}] ratios are shown. For the purposes of this test,
    the Eu abundance was set to the upper limit value.

The results in Table \ref{table:SysErrors} demonstrate that the
[\ion{Fe}{1}/H] ratio is relatively insensitive ($<0.1$ dex) to the
isochrone parameters,\footnote{This insensitivity to the isochrone
  parameters makes it possible to use [\ion{Fe}{1}/H] as a constraint
  on an appropriate isochrone age and metallicity, as in Figures
  \ref{fig:AllTrends} and \ref{fig:FeHOffset}.} while \ion{Fe}{2}
shows a strong sensitivity to isochrone metallicity.  Most of the
[X/Fe] ratios are insensitive to age shifts of 4 Gyr,  especially
when the age is increased.  Lowering the age to 6 Gyr does moderately
affect many of the singly ionized species, but only when
[X/\ion{Fe}{2}] ratios are used.  Changes in the isochrone metallicity
have a much stronger effect on all abundance ratios other than
[Mg/Fe], [Ni/Fe], and [Cu/Fe].  In particular, the Ti, Zn, Y, Ba, and
Eu abundance ratios can be affected by more than 0.1 dex. Changing the
[$\alpha$/Fe] of the isochrone has a negligible effect on all
abundance ratios.

\begin{table}
\centering
\begin{center}
\caption{Systematic uncertainties based on uncertainties in
  isochrone properties.\label{table:SysErrors}}
  \newcolumntype{e}[1]{D{.}{.}{#1}}
  \newcolumntype{d}[1]{D{,}{\;\pm\;}{#1}}
  \begin{tabular}{@{}le{3}e{3}e{3}e{3}e{5}@{}}
  \hline
 & \multicolumn{2}{c}{$\Delta$ Age (Gyr)} & \multicolumn{2}{c}{$\Delta$ [Fe/H]} & \multicolumn{1}{c}{$\Delta$ [$\alpha$/Fe]}\\
   & \multicolumn{1}{c}{$-4$} & \multicolumn{1}{c}{$+4$} &
  \multicolumn{1}{c}{$-0.3$} & \multicolumn{1}{c}{$+0.3$} & \multicolumn{1}{c}{$-0.4$} \\ 
  \hline
$\Delta[$\ion{Fe}{1}/H$]$ & +0.08 & -0.04 & +0.05 &  0.0  & -0.03 \\
$\Delta[$\ion{Fe}{2}/H$]$ & -0.01 &  0.0  & -0.16 & +0.23 & +0.03 \\
$\Delta[$Na/Fe$]$ & -0.02 &  0.0  & +0.03 & -0.07 &  0.0  \\
$\Delta[$Mg/Fe$]$ &  0.0  &  0.0  & +0.01 & -0.02 & +0.01 \\
$\Delta[$Al/Fe$]$ & -0.03 & +0.01 & +0.03 & -0.06 & 0.0 \\
$\Delta[$Ca/Fe$]$ & +0.01 & -0.01 & +0.06 & -0.08 & -0.01 \\
$\Delta[$\ion{Ti}{1}/\ion{Fe}{1}$]$ & +0.04 & -0.03 & +0.13 & -0.13 & -0.03 \\
$\Delta[$\ion{Ti}{2}/\ion{Fe}{1}$]$ & -0.03 &  0.0  & -0.11 & +0.14 & +0.03 \\
$\Delta[$\ion{Ti}{2}/\ion{Fe}{2}$]$ & +0.06 & -0.04 & +0.10 & -0.09 & -0.03 \\
$\Delta[$Cr/Fe$]$ & +0.03 & -0.02 & +0.08 & -0.11 & -0.02 \\
$\Delta[$Mn/Fe$]$ & +0.01 &  0.0  & +0.05 & -0.06 & -0.01 \\
$\Delta[$Ni/Fe$]$ & +0.01 & -0.01 & +0.01 & +0.02 &  0.0 \\
$\Delta[$Cu/Fe$]$ & -0.01 & +0.01 & +0.02 &  0.0  &  0.0 \\
$\Delta[$Zn/Fe$]$ & -0.02 & +0.01 & -0.08 & +0.10 & +0.03 \\
$\Delta[$\ion{Y}{2}/\ion{Fe}{1}$]$  & -0.02 & -0.01 & -0.11 & +0.11 & +0.03 \\
$\Delta[$\ion{Y}{2}/\ion{Fe}{2}$]$  & +0.07 & -0.05 & +0.10 & -0.12 & -0.03 \\
$\Delta[$\ion{Ba}{2}/\ion{Fe}{1}$]$ & -0.01 &  0.0  & -0.11 & +0.12 & +0.02 \\
$\Delta[$\ion{Ba}{2}/\ion{Fe}{2}$]$ & +0.08 & -0.04 & +0.10 & -0.11 & -0.04 \\
$\Delta[$\ion{Eu}{2}/\ion{Fe}{1}$]$ & -0.05 & +0.02 & -0.15 & +0.16 & +0.04 \\
$\Delta[$\ion{Eu}{2}/\ion{Fe}{2}$]$ & +0.04 & -0.04 & +0.06 & -0.07 & -0.02 \\
\hline
\end{tabular}
\end{center}
\medskip
\raggedright .\\
\end{table}

From the results in Table \ref{table:SysErrors}, it is unclear whether
[X/\ion{Fe}{1}] or [X/\ion{Fe}{2}] ratios are more robust for singly
ionized species.  In the case of \ion{Ti}{2}, comparisons with
\ion{Fe}{1} lead to lower systematic errors for age shifts, while
\ion{Fe}{2} leads to lower offsets for [Fe/H] shifts.  For \ion{Y}{2}
and \ion{Ba}{2} the differences between the \ion{Fe}{1} and
\ion{Fe}{2} ratios are minimal for metallicity shifts, while
\ion{Fe}{1} is preferable for age shifts.  Finally, for \ion{Eu}{2}
the \ion{Fe}{2} ratios are more robust to age and metallicity shifts.
Ultimately, there is a lot of complexity behind these abundance
ratios, related to the specific spectral lines and their properties.
However, because it is not immediately clear which Fe ionization state
to use for singly ionized species, both are reported throughout this
paper.


\section{Discussion}\label{sec:Discussion}
As mentioned in Section \ref{sec:Intro}, G1 has long been identified
as both a GC and a possible former NSC.  Below these possibilities are
discussed in the context of G1's detailed abundance ratios.

\subsection{G1 as a Globular Cluster}\label{subsec:G1GC}
Although G1 physically resembles a GC, there are many open
questions as to whether it can truly be considered a GC and whether GC
trends can be scaled up to G1's mass.  This section discusses how G1's
detailed abundances can shed light on this issue.

\subsubsection{Sodium and Aluminum enhancement}\label{subsubsec:Na}
One of the most striking results from this abundance analysis of G1 is
the elevated [Na/Fe] and [Al/Fe] ratios, at $+0.6$ and $+0.72$ dex,
respectively.  A Na-O anticorrelation is a ubiquitous feature of
classical Milky Way GCs, while many clusters have also found to
exhibit a Mg-Al anticorrelation (e.g.,
\citealt{Carretta2009}). Numerous IL studies of Milky Way and
extragalactic GCs have found that many clusters have elevated IL
[Na/Fe] ratios and, in some cases, elevated [Al/Fe] ratios, which may
be signatures of multiple populations (e.g.,
\citealt{Colucci2009,Colucci2014,Colucci2017},
\citealt{Sakari2013,Sakari2015,Sakari2016},
\citealt{Larsen2017,Larsen2018a,Larsen2018}). G1's elevated [Na/Fe]
and [Al/Fe] seem to indicate that it too hosts Na- and Al-enhanced
stars---the tests in Section \ref{subsec:SysErrors} have demonstrated
that elevated Na and Al cannot be caused by systematic errors in
isochrone selection. Since elevated [Na/Fe] and [Al/Fe] are a unique
signature of GCs, this finding indicates that G1 is intimately
connected with the classical GCs.

\begin{figure*}
\begin{center}
\centering
\includegraphics[scale=0.75,trim=0.0in 0 0.0in 0.0in,clip]{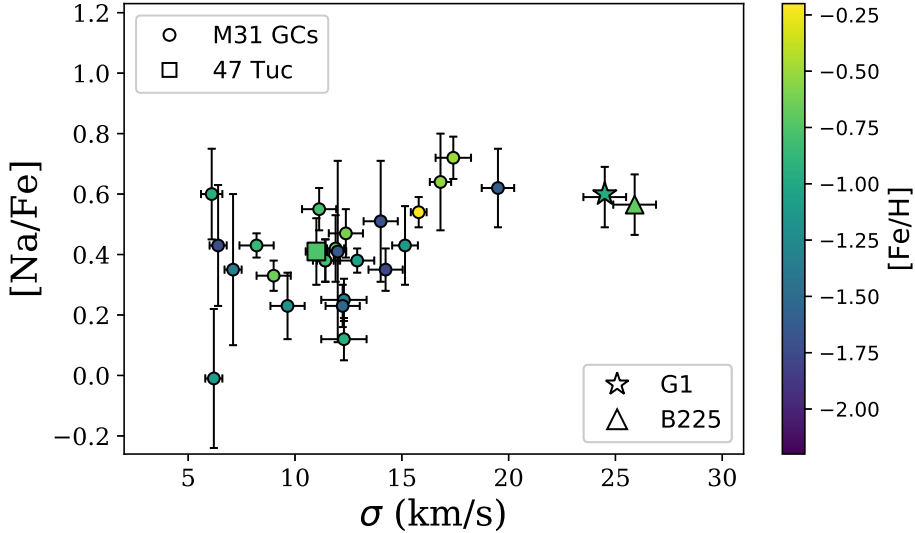}
\caption{The IL [Na/Fe] ratios in M31 GCs as a function of velocity
  dispersion.  The star shows G1, the square shows 47~Tuc (this
  analysis), the circles show other M31 GCs (\citealt{Colucci2014} and
  \citealt{Sakari2015,Sakari2016}), and the triangle shows B225
  \citep{Larsen2018}.  The points  are color-coded by cluster
        [Fe/H].}\label{fig:NaMass}
\end{center}
\end{figure*}

G1's [Na/Fe] and [Al/Fe] ratios are also $\sim0.2$ and $\sim0.4$ dex
higher than 47~Tuc's value, respectively.  One possible interpretation
of this enhancement is that G1 may possess a higher fraction of
Na-enhanced stars (either I or E stars in the \citealt{Carretta2010}
framework) than 47~Tuc.  G1 may have relatively more Na-enhanced stars
as a result of its  higher mass: Figure \ref{fig:NaMass} shows the IL
[Na/Fe] versus the cluster velocity dispersion, a proxy for cluster
mass,\footnote{Note that the velocity dispersion reflects the
  current cluster mass, which may differ from its birth mass.} for a
sample of M31 GCs; 47~Tuc is also shown for comparison. This figure
demonstrates that the higher mass, higher velocity dispersion clusters
tend to have  higher IL [Na/Fe] ratios. G1's high [Na/Fe] is very
similar to B225, another massive cluster \citep{Larsen2018}.  This
trend with mass is also reinforced by observations of Milky Way GCs,
where more massive clusters seem to have relatively fewer of the
``primordial'' population stars \citep{Milone2017}.  Similar trends
with mass (or proxies for mass) have also been previously detected in
IL [Na/Fe] ratios \citep{Colucci2014}, [Na/O] ratios
\citep{Sakari2016}, and [N/Fe] ratios \citep{Schiavon2013}.  A general
interpretation of these trends with mass is that more massive clusters
are able to retain more ejecta from the source of multiple
populations, leading to an increased number of Na- and N-enhanced
stars (see the review by \citealt{BastianLardo2018}).  This result may
extend to Al, indicating that G1 also has an intracluster spread
in Al, similar to several Milky Way GCs (e.g.,
\citealt{Carretta2009}).\footnote{Note that although 47~Tuc's stars
  are Al-enhanced, they may be consistent with standard Milky Way
  thick disk trends, with no intracluster
  spreads. \citet{Thygesen2014} find that an apparent Al spread could
  occur soley because of NLTE effects.  \citet{Thygesen2016} further
  find no intracluster changes in Mg isotopes, suggesting that Mg-Al
  cycling has not occurred in 47~Tuc.  \citet{Meszaros2020}
  additionally find no evidence for an [Al/Fe] spread within 47~Tuc.}

Follow-up spectroscopy further in the blue or in the infrared may
better characterize the multiple populations in G1.  Based on infrared
IL spectra of other M31 GCs, \citet{Sakari2016} were able to obtain O
abundances and found a significant trend in [Na/O] versus cluster
mass; however, O abundances cannot be determined from G1's optical
spectrum. Similarly, a N abundance cannot be determined with the
spectrum in this paper, meaning that G1 cannot yet be compared to
the trend detected by \citet{Schiavon2013}.  Despite its high [Al/Fe],
it is worth noting that G1's [Mg/Fe] agrees with [Ca/Fe] and
[\ion{Ti}{1}/Fe], suggesting that there is not a significant Mg spread
in the cluster.  However, this is consistent with the individual
stellar abundances in Milky Way GCs; \citet{Carretta2009} find that in
many clusters a fairly large spread in [Al/Fe] can be accompanied by a
much smaller spread in [Mg/Fe].  Follow-up of stronger Al features may
better characterize the nature of the multiple-populations in G1.

To summarize, G1's elevated [Na/Fe] and [Al/Fe] suggest that it is
intimately connected with GCs.  Even if G1 is considered to be a NSC
rather than a classical GC, it seems to have shared a common formation
pathway with the classical GCs (see Section \ref{subsubsec:NSCs}).

\subsubsection{Fe Spreads}\label{subsubsec:TypeIIGCs}
As mentioned throughout this paper, G1 has photometric evidence for
intracluster Fe and He spreads \citep{Meylan2001,Nardiello2019}.  The
presence of an iron or helium spread does not automatically disqualify
G1 as a GC.  There are at least ten Milky Way GCs with broadened or
bifurcated RGBs (the so-called ``Type II'' GCs; \citealt{Milone2017});
several of these GCs, notably $\omega$ Cen, have been confirmed to be
iron-complex GCs (e.g., \citealt{JohnsonPilachowski2010}).
\citet{Milone2017} and \citet{Marino2019} also note that these Type II
GCs can show spreads in He, CNO, and $s$-process elements.  A future
paper (Sakari et al. {\it in prep.}) will investigate the effects of
abundance spreads on the IL abundances, including whether such spreads
could be detected in an IL spectrum.  However, it is worth considering
how G1's IL values agree with those of the Type-II GCs.

Amongst the population of Type-II Milky Way GCs, G1 falls on the
massive and metal-rich end.  Of the Type II clusters in
\citet{Milone2017}, only NGC~6388 has an average [Fe/H] as high as G1
\citep{Harris}---the rest are predominantly more metal-poor.
\citet{Marino2019} find that for Milky Way GCs the $\Delta$[Fe/H]
should increase with cluster mass; extrapolating the Milky Way results
in their Figure 20 shows that G1 could have $\Delta[\rm{Fe/H}]\ga
0.4$ dex, a spread that disagrees with the photometric analysis by
\citet{Nardiello2019}.  One possibility for this disagreement is that
the {\it HST} photometry of the outer regions misses part of G1's
stellar population.  Future follow-up photometry, e.g., in the crowded
central regions, could help to characterize the metallicity
distribution function in G1.

\subsubsection{Neutron Capture Abundances}\label{subsubsec:NeutronCaptureDiscussion}
The abundances in Table \ref{table:Abunds} show that G1 has a solar
[Ba/Fe] ratio, slightly subsolar [Y/Fe], and a moderate upper limit in
[Eu/Fe], indicating that it is not significantly enhanced in neutron
capture elements.  These values are similar to those of 47~Tuc and
other classical Milky Way GCs at G1's metallicity.  This lack of
enhancement is distinct from \omgCennosp, whose intermediate metallicity
stars have supersolar [La/Fe] ratios
\citep{JohnsonPilachowski2010};\footnote{La is primarily an
  $s$-process element, like Ba \citep{Burris2000}.  Because its lines
  are weaker, La is often more desirable for individual stellar
  analyses; however, these weaker lines are not detectable in G1.}
the stars in \omgCen show a clear rise in [La/Fe] with increasing
metallicity above $[\rm{Fe/H}]\sim-1.7$
\citep{JohnsonPilachowski2010,Marino2011}.  Although this increasing
trend may flatten in the most metal-rich stars, most of \omgCennosp's
metal-rich stars are still enhanced in La
\citep{JohnsonPilachowski2010}.  The low IL [Ba/Fe] for G1 suggests
that G1's stars are not significantly enhanced in Ba, and therefore
that G1 has not experienced the same ongoing enrichment in
neutron-capture elements (though see Sakari et al., {\it in prep.},
for tests of abundance spreads).

G1's lower limit in [Ba/Eu] shows that the cluster has experienced at
least some small enrichment in $s$-process elements.  Again, this
level of $s$-process enrichment is consistent with the chemical
evolution pattern of typical Milky Way field stars and GCs like
47~Tuc.  However, G1 is again discrepant from \omgCennosp, which has
supersolar enhancement in [La/Eu].  A high [La/Eu] ratio indicates
that the high [La/Fe] ratios in \omgCen are the result of the
$s$-process rather than the $r$-process.  Since this analysis can only
provide a lower limit in [Ba/Eu] for G1, it is possible that the
cluster could have had more contributions from the $s$-process, but
this seems unlikely based on its solar [Ba/Fe] ratio.

\citet{DOrazi2011} also note that \omgCennosp's metal-rich stars have
supersolar [Ba/Y] ratios, which indicates an excess of lighter
$s$-process elements like Y.  They argue that the low [Ba/Y] in
\omgCen is a signature of extra contributions from the higher mass AGB
stars. Since G1's [Ba/Y] ratio is supersolar (again, in agreement with
47~Tuc), it seems that G1 does have a slight enhancement in lighter
$s$-process elements.  This [Ba/Y] ratio will be discussed further in
Section \ref{subsubsec:DwarfAlpha}.

Altogether, G1's IL neutron-capture element ratios are similar to
47~Tuc and other classical GCs.  Unlike \omgCennosp, G1's neutron capture
abundances agree with those expected for a standard Milky Way cluster
at $[\rm{Fe/H}] = -1$.  Of course, the stars in \omgCen have been
studied individually, not via IL, which makes it easier to
characterize the intracluster trends.  The second paper in this
series, Sakari et al. ({\it in prep.}), will examine the effects of
intracluster abundance spreads on the IL spectrum.

\subsubsection{Comparisons with M31 GCs}\label{subsubsec:M31GCs}
Compared with Milky Way GCs, G1 is unusual because it is is more
massive and more metal-rich than typical Milky Way GCs.  M31, however,
contains multiple, massive, G1-like clusters.  A handful of these GCs
(B023, B158, and B225) have been photometrically identified as
candidate iron-complex clusters, all of which were found to have
$[\rm{Fe/H}]~\sim~-0.9$ to $-0.6$ \citep{FuentesCarrera2008}.
\citet{Colucci2014}, \citet{Sakari2016}, and \citet{Larsen2018} have
all conducted high-resolution IL spectroscopic analyses of B225,
finding $[\rm{Fe/H}]\sim-0.6$. \citet{Sakari2016} and
\citet{Larsen2018} further found minimal offsets between the optical
and infrared [Fe/H] ratios, placing constraints on the extent of an
iron spread.  Another massive M31 GC, B088, is more metal-poor
($[\rm{Fe/H}]=-1.7$; \citealt{Sakari2016}), similar to $\omega$~Cen;
\citet{Sakari2016} suggested that variations in [Mg/Fe] between the
optical and infrared could indicate an iron spread within B088.  There
are several additional massive clusters that have not been studied in
any detail.  Ultimately, G1 is not unique in being a massive,
moderately metal-poor star cluster with a possible iron spread.  While
it is the only one of these iron-complex clusters that is obviously in
the outer halo \citep{Mackey2019}, \citet{Perina2012} argue that B023
and B225 are actually outer-halo clusters that are projected into the
inner regions.

Other than its enhanced [Na/Fe], G1 seems chemically similar to other
M31 GCs.  Unlike the Milky Way GCs, which exhibit a bimodal [Fe/H]
distribution (e.g., \citealt{FreemanNorris1981}), the M31 GCs don't
have a clear [Fe/H] bimodality \citep{Caldwell2011}. M31 also contains
more metal-rich GCs than the Milky Way: the median [Fe/H] of the
\citet{Caldwell2011} sample is $[\rm{Fe/H}]~=~-0.90\pm0.07$, compared
to a median $[\rm{Fe/H}]~=~-1.35\pm0.14$ in the Milky Way
\citep{Harris}.  G1 therefore has a typical metallicity for an M31 GC.
Although the outer halo GCs are, in general, more metal-poor than
those in the inner regions (e.g., \citealt{Caldwell2011}), there are
multiple metal-rich GCs that are projected into the outer-halo (e.g.,
\citealt{Sakari2015}).  G1's [$\alpha$/Fe] enhancement is also in
agreement with other M31 GCs at $[\rm{Fe/H}]\sim-1$ in the inner and
outer halo (see Figure \ref{fig:CaFe}).

G1's properties are therefore consistent with those of other M31 GCs,
including those in the outer halo.  Its mass-to-light ratio is also
similar to other M31 GCs \citep{Strader2009}. If it were not for its
high mass and high [Na/Fe], G1 would be considered a typical M31 GC
based on its IL abundances.  However, G1's high mass and its presence
in the outer halo do raise some questions about where and how it
formed.  Many other open questions also remain about the origin of the
outer halo and its GCs in general (e.g.,
\citealt{McConnachie2018,Mackey2019}). G1's relationship to the outer
halo of M31 will be discussed further in Section
\ref{subsubsec:M31OH}. 

\subsection{G1's Potential Origins in a Dwarf Galaxy}\label{subsec:G1NSC}
Because of its mass and putative Fe spread, G1 has long been
identified as the potential former NSC of a dwarf galaxy, like
$\omega$ Cen. Under this paradigm, G1 would have assembled at the
center of a satellite dwarf through GC mergers, {\it in situ} star
formation, or a combination of the two processes (see
\citealt{Neumayer2020}).  As this dwarf galaxy fell into M31's
gravitational potential, the outer field stars (and other GCs, if
present) in the dwarf galaxy would have been stripped, creating
stellar streams (see, e.g., simulations by \citealt{BekkiChiba2004}).
Even if G1 was not a NSC,  its location in the outer halo
($R_{\rm{proj}} = 34.7$ kpc; \citealt{Mackey2019}) suggests that it
may have originated as a GC in a dwarf galaxy.  This section discusses
G1's origins in light of its detailed abundances.

\subsubsection{Connections with Known Nuclear Star
  Clusters}\label{subsubsec:NSCs}
\cms{NSCs are found in many types of galaxies, from dwarf ellipticals
  to massive spirals like the Milky Way. The NSCs themselves have a
wide range of properties, varying with host galaxy type, galaxy mass,
NSC mass, etc.  (see, e.g., \citealt{Neumayer2020}).  This discussion
focuses specifically on the NSCs that are associated with dwarf
galaxies; however, it is worth noting that the properties of NSCs can
still vary significantly, even in dwarf galaxies.}

Past photometric results showed that NSCs in dwarf galaxies have
similar colors as the Milky Way GCs which have extremely blue
horizontal branches (including $\omega$ Cen), \cms{which suggests that
dwarf galaxy NSCs and GCs} have similar ages and metallicities
\citep{Georgiev2009}.  \citet{OrdenesBriceno2018} similarly found
dwarf nuclei to have colors consistent with metal-poor stellar
populations. Using a variety of techniques, several medium-resolution
IL spectroscopic studies of NSCs in dwarfs have found varying
metallicities.  \citet{Spengler2017} performed Lick index analyses on
12 confirmed or candidate dwarf elliptical NSCs in the Virgo galaxy
cluster, finding ages $\sim2-11$ Gyr and $[\rm{Fe/H}]\sim-1.2$ to
$-0.15$.  \citet{Kacharov2018} and \citet{Fahrion2020} used full
spectrum fitting with different techniques to model the stellar
populations in NSCs that are associated with a variety of galaxy
types.  For a single NSF in a  dwarf elliptical/S0 galaxy NSC,
\citet{Kacharov2018} found an age pf $\sim0.5-3$ Gyr and
$[\rm{Fe/H}]\sim-0.4$ to $-0.2$, depending on the fitting technique
used.  \citet{Fahrion2020} found the nuclei of two dwarf satellites of
Cen~A to be older ($\sim7$ Gyr) and more metal-poor
($[\rm{Fe/H}]\sim-1.8$).  At $[\rm{Fe/H}]=-0.98$, G1 certainly falls
within the observed metallicity ranges of NSCs.  Though its age of
10~Gyr is older than most of the estimates for other NSCs, it is
important to remember that these ages were derived with different
techniques and that it is notoriously difficult to obtain ages from IL
spectra (as discussed in \citealt{Fahrion2020}).\footnote{Note that
  \citet{Spengler2017} find older ages for many of their NSCs when
  photometric spectral energy distributions are used.}

It is also worth considering whether G1 agrees with the trends from
known NSCs.  Amongst populations of NSCs in early and late-type
galaxies, G1's velocity dispersion and mass-to-light ratio falls
within the known ranges (e.g., \citealt{Boker2004,Leigh2012}).
\citet{Spengler2017} and \citet{SanchezJanssen2019} both
present empirical relationships between the mass of a NSC and the
stellar mass of its host galaxy.  For G1, with a mass of
$\sim10^{7.2}$~M$_{\sun}$ \citep{Meylan2001,Nardiello2019}, these
empirical trends would suggest that G1 originated in a galaxy with a
stellar mass $M_{\star}~=~10^8-10^{10}$~M$_{\sun}$, i.e., about the
mass of the Magellanic clouds \citep{Kim1998}. \cms{
\citet{Neumayer2020} also present a compilation of galaxy masses and
NSC metallicities from the literature.  This relation shows that
G1's [Fe/H] is consistent with origins in a galaxy with a stellar
mass $\sim 10^9$~M$_{\sun}$---note that \citet{Neumayer2020} argue
that NSC formation pathways change at this galaxy mass.  They
predict that NSCs associated with lower mass galaxies form primarily
through GC mergers, while those associated with more massive
galaxies form primarily through ongoing {\it in situ} star
formation. G1's enhanced [Na/Fe] indicates that, if it is a NSC, it
has an intimate connection with GCs.}

It is also worth noting that G1 does not appear to contain stars as
metal-poor as M54, the NSC of the Sagittarius (Sgr) dwarf spheroidal
\citep{Mucciarelli2017}.  A lack of metal-poor stars could also
suggest that if G1 is a NSC, then it originated in a galaxy more
massive than Sgr, \cms{which is estimated to have had a total stellar
  mass $\sim10^8$ M$_{\sun}$
  \citep{VasilievBelokurov2020}.\footnote{\cms{Note that current mass
    estimates of the Sgr stream are lower than than the total mass
    estimate from \citet{VasilievBelokurov2020} because the galaxy is
    being tidally disrupted.  \citet{Penarrubia2011} find that as much
    as 40-50\% of the initial stellar mass may have been lost.}}}  More
work should be done to understand whether G1 truly fits in with the
population of NSCs.

\subsubsection{Clues from Chemical Abundances}\label{subsubsec:DwarfAlpha}
\begin{figure*}
\begin{center}
\centering
\includegraphics[scale=0.75,trim=0in 0 0.0in 0.0in,clip]{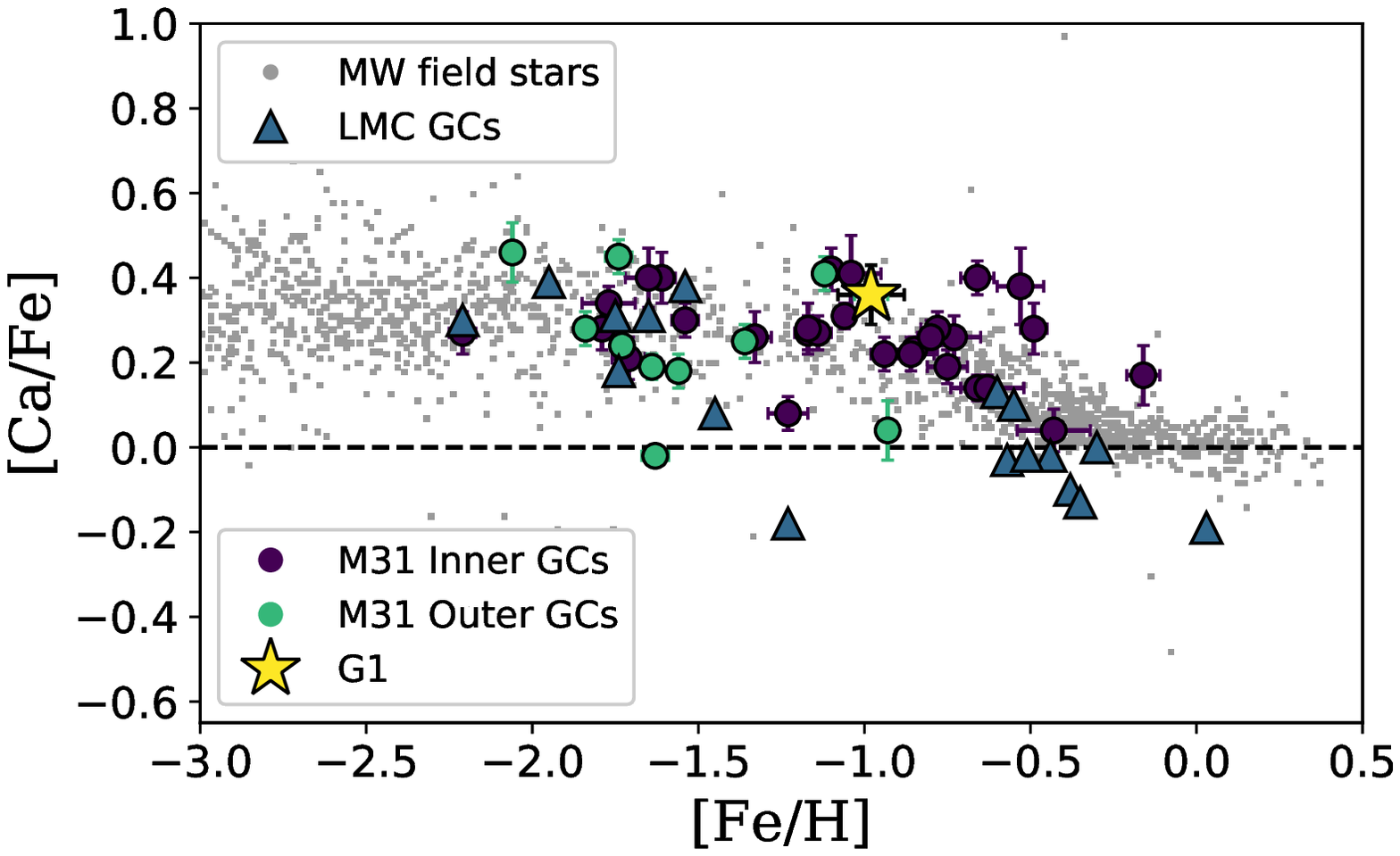}
\caption{[Ca/Fe] versus [Fe/H] in G1 (yellow star) compared with Milky
  Way field stars (grey points;
  \citealt{Venn2004,Reddy2006,Sakari2018}), LMC GCs (IL and averages
  of individual stars; triangles,
  \citealt{Johnson2006,Mucciarelli2008,Mucciarelli2010,Colucci2012,Sakari2017}),
    and M31 inner and outer GCs (IL abundances; circles,
    \citealt{Colucci2014,Sakari2015,Sakari2016}).}\label{fig:CaFe}
\end{center}
\end{figure*}

G1 shows enhanced [$\alpha$/Fe], based on its [Mg/Fe], [Ca/Fe], and
[Ti/Fe] ratios.  Figure \ref{fig:CaFe} compares G1's IL [Ca/Fe]
vs. [Fe/H] ratios to field stars in the Milky Way and GCs in the LMC
and M31.  Plots such as this are often used to unravel a galaxy's
chemical evolution history, specifically the onset of Type Ia
supernovae which leads to a downturn in [Ca/Fe] with increasing [Fe/H]
(see, e.g., \citealt{MatteucciBrocato1990,Tolstoy2009}).  Several papers have used the
[$\alpha$/Fe] ratios to identify GCs that may have been accreted from
low-mass dwarf galaxies.  In particular, the three metal-poor GCs in
Figure \ref{fig:CaFe} with lower [Ca/Fe] than the other M31 GCs (G002,
B457, and PA-17) have been identified as candidate accreted GCs
\citep{Colucci2014,Sakari2015,Sakari2016}.  G1's high [Ca/Fe] at
$[\rm{Fe/H}]~=~-0.98$ indicates that it could not have originated in a
very low-mass dwarf spheroidal.

\citet{McWilliam2013} note that Cu is another useful element for
chemical tagging.  In their analysis of three stars in the Sgr dwarf
spheroidal, \citet{McWilliam2013} found lower [Cu/Fe] compared to
Milky Way field stars.  However, the [Cu/O] ratio fit the general
trend with [Fe/H] seen in field stars, consistent with
metallicity-dependent Cu production by massive stars (e.g.,
\citealt{Pignatari2010}).  G1's [Cu/Fe] ratio, although low, is
consistent with the Cu abundances of typical Milky Way field stars at
$[\rm{Fe/H}] = -1$; 47~Tuc's higher [Cu/Fe] ratio, compared to G1, can
be explained by normal chemical evolution.  G1's [Cu/Fe] ratio is
therefore inconsistent with formation in a very low-mass galaxy.

In Sgr dwarf spheroidal stars, \citet{McWilliam2013} also found low
[Mg/Ca] ratios, which they attributed to a top-light IMF, i.e., a lack
of the most massive stars, within Sgr.  They argued that a top-light
IMF could be a natural outcome in a low-mass system which might be
unable to form the large giant molecular clouds that are needed to
create the highest mass stars.  G1's normal [Mg/Ca] ratio requires no
modifications to the IMF.

Finally, supersolar [Ba/Y] (or [La/Y]) has also been identified as a
chemical signature of stars that form in low-mass dwarf galaxies
(e.g., \citealt{Shetrone2003,Letarte2010,Sakari2011}).  One
explanation for high [Ba/Y] ratios is that the $s$-process elements in
dwarf galaxies are created in lower-metallicity AGB stars; as a result
of their lower metallicity, these AGB stars have fewer seed nuclei,
and can build up their $s$-process elements to higher proton numbers.
G1 shows only a moderate enhancement in [Ba/Y], which is again
consistent with 47~Tuc and other Milky Way field stars and clusters.
It therefore seems that G1's $s$-process enhancement did not come from
AGB stars that were as metal-poor as those in very low-mass dwarf
galaxies.

Altogether, G1's abundances are more consistent with origins in a
fairly massive galaxy, perhaps one that was at least as massive as the
Large Magellanic Cloud (LMC), which has a stellar mass
$\sim10^9$~M$_{\sun}$ \citep{Kim1998}.  It is worth noting that the
presence of $s$-process elements requires contributions from AGB
stars, while the elevated $[\alpha$/Fe] ratios rules out contributions
from Type Ia supernovae.  \cms{This places important constraints on the
timescales for G1's formation: AGB star evolution requires timescales
$\sim1$~Gyr (e.g., \citealt{BaSTIREF}) while Type Ia supernovae start
occurring on a similar timescale \citep{Maoz2010}.}  Future follow-up
observations of, e.g., Rb and Zr could further constrain the mass of
the AGB progenitors that created the $s$-process material in G1.

\subsubsection{Connections with M31's Outer Halo}\label{subsubsec:M31OH}
One of the intriguing characteristics of G1 is that though it has long
been identified as a potential accreted cluster, it has no clear links
with other GCs, intact dwarfs, or bright stellar streams.  In
projection, G1 does seem to lie near a group of GCs known as
``Association 2''---however, G1's discrepant radial velocity suggests
that it is not actually physically associated with those GCs
\citep{Veljanoski2014}.  G1 is also projected close to a stellar
over-density known as the ``G1 clump'' (so-named due to its proximity
to G1; \citealt{Ferguson2002}), although these stars seem to be from
M31's disk \citep{Ferguson2005,Faria2007,Richardson2008}.
From spectra of individual stars in a region near G1 and the G1 clump,
\citet{Reitzel2004} found that only a handful of stars had radial
velocities consistent with G1, indicating that G1 likely does not have
bright tidal tails.  There is also no photometric evidence of stellar
debris surrounding G1 (e.g, \citealt{Mackey2019}).\footnote{Note that
  \citet{Mackey2019} used a ``density percentile value'' to
  quantify the density of stars surrounding outer halo GCs; they
  subsequently used this value to identify clusters that may be
  associated with faint substructure.  Although G1 has a low density
  percentile value, indicating an absence of bright substructure,
  \citet{Mackey2019} classified G1 as a cluster that is associated
  with substructure, based on its possible classification as a NSC.}
The question then arises: if G1 {\it did} originate in a dwarf galaxy,
where is the remainder of that galaxy now?

One potential explanation for a lack of bright streams around G1 is
that the host galaxy was accreted early on, so that any streams have
since dissolved (e.g., \citealt{Johnston2008}). \citet{Ibata2014} and
\citet{Mackey2019} argue that early accretion can explain the lack of
substructure in the most metal-poor outer halo stars and some of the GCs.
However, outer halo stars with metallicities as high as G1 are
primarily associated with substructure, particularly the Giant Stellar
Stream to the south of M31 \citep{Ibata2014}.  The lack of a
``smooth'' metal-rich component in the outer halo may indicate that a
massive, metal-rich, LMC-like galaxy could not have been accreted very
early on.  This discrepancy may worsen if G1 was a
NSC---\citet{Johnston2020} find evidence that NSCs in the Fornax
cluster are generally more metal-poor and $\alpha$-rich than the field
stars in their main galaxy.  If G1's host contained stars more
metal-rich than $[\rm{Fe/H}]>-0.7$, it is not clear where those stars
are now.  Alternatively, a lack of nearby streams could indicate that
the NSC was a dominant part of the galaxy's total mass.
\citet{SanchezJanssen2019} find that in low-mass galaxies a NSC can
comprise up to $\sim50$\% of the galaxy's total stellar mass.  Of
course, comparisons with more isolated, intact nucleated dwarfs may
not be appropriate, since G1's host galaxy may have been disturbed
during its early evolution.  Ultimately, deeper imaging of the
vicinity around G1 could help assess the presence of stellar streams
around the cluster.

Intuitively, G1's lack of associated GCs seems problematic, since it
is unusual for massive galaxies to have only a single, massive,
metal-rich GC; such a galaxy would likely have a very low specific
frequency (e.g., \citealt{Harris2013}).  However, if G1 was a NSC it
may have no associated GCs.  Based on observations of NSCs in the
Virgo cluster, \citet{SanchezJanssen2019} find that though GCs or NSCs
are similarly common in galaxies (i.e., the two objects similar
``occupation fractions''), the number of galaxies with both NSCs and
GCs is slightly lower.  In this scenario, G1's host galaxy may not
have possessed any other GCs, which explains why none can be
kinematically and chemically linked to G1 now.\footnote{Note that G1
  is kinematically and spatially close to the cluster G002---however,
  G002 is much more metal-poor ($[\rm{Fe/H}]~=~-~1.63$) and
  $\alpha$-poor ($[\rm{Ca/Fe}]~=~-0.02$; \citealt{Colucci2014}) than
  G1.  It is difficult to imagine a scenario where the two could be
  associated without invoking complicated chemical evolution
  scenarios.}  It is more unusual, however, that G1's progenitor would
not have contained any metal-poor GCs, since most galaxies, especially
dwarfs, have a significant population of metal-poor GCs (see, e.g.,
\citealt{BrodieStrader2006}; several metal-poor LMC clusters are also
visible in Figure \ref{fig:CaFe}).  More modelling should be done to
see if G1 can be kinematically linked to any other GCs or known
stellar streams.

It is worth noting that other than its high [Na/Fe], G1 chemically
resembles the classical M31 GCs, the Milky Way GCs (e.g., 47~Tuc), and
the Milky Way disk stars.  \cms{One simple explanation for this
  chemical similarity in the different environments is that G1 formed
  in M31 itself, rather than in a dwarf satellite.}  Its proximity to
the G1 clump, material that is believed to have originated in M31's
disk, further hints that G1 could have been born out of material
from the disk. Based on observations of lower mass clusters in M31's
disk, \citet{Johnson2017} argue that the high-mass limit of the
cluster mass function depends on a galaxy's star formation rate
surface density (i.e., the intensity of local star formation).  The
formation of a massive cluster like G1 through {\it in situ} star
formation alone would require a very high rate of star formation. One
could perhaps envision a scenario that would lead to the formation of
a massive cluster in the disk, e.g., an interaction with a very high
mass satellite galaxy---however, it is unclear how a massive cluster
like G1 could be removed from the disk and brought into the outer
halo.

Finally, it is worth noting that chemically G1 is also very similar to
B225, the other massive cluster that has been studied at high spectral
resolution, except that B225 is slightly more metal-rich
\citep{Colucci2014,Sakari2016,Larsen2018}.  There are several
additional unstudied, massive GCs in the inner regions, many of which
appear to be similarly metal-rich.  Any formation
scenario for G1 should be able to explain the properties of these
other GCs as well.  Given that G1 seems to lack metal-poor stars (see
the photometric limits on $\Delta$[Fe/H] by \citealt{Nardiello2019}),
a GC merger scenario would have to serendipitously involve GCs of
roughly the same moderate metallicity. Such a scenario seems unlikely
for all the massive, metal-rich GCs in M31's halo.  Follow-up
observations of these other clusters, specifically to characterize the
intracluster iron spreads, would be useful for interpreting the results
in G1.

Ultimately the abundances derived in this paper cannot shed light on
whether G1 originated in a dwarf galaxy or not, though they do
indicate that its birth site was a fairly massive galaxy, at least as
massive as the LMC.

\section{Conclusion}\label{sec:Conclusion}
This paper has presented a high-resolution, IL spectroscopic abundance
analysis of the massive M31 cluster, G1.  A future paper will explore
the effects of intracluster abundance spreads on the IL abundances.
When a single age and metallicity is adopted for the entire cluster,
G1 was found to be old, with an age of 10~Gyr, and moderately
metal-poor, at $[\rm{Fe/H}]=-0.98\pm0.05$.  The cluster has $\alpha$,
[Cu/Fe], \cms{[Mg/Ca]}, Fe-peak and neutron-capture element abundance ratios
that are typical for Milky Way and M31 GCs at $[\rm{Fe/H}] = -1$.
This comparison is strengthened by the similarity between G1 and the
Milky Way cluster 47~Tuc.  These abundances place constraints on G1's
birth environment, suggesting that G1 formed in a galaxy that was at
least as massive as the LMC.

G1 was also found to have elevated Na and Al, with ratios
$[\rm{Na/Fe}]~=~+0.60$ and $[\rm{Al/Fe}]~=~+0.72$, suggesting that it
contains a high fraction of Na- and Al-enhanced (and, presumably,
O-deficient) stars, a unique chemical signature of GCs.  This result
indicates that G1 shared a similar formation pathway as GCs.  If G1 is
a former NSC, it could have formed through {\it in situ} star
formation, GC mergers, or a combination of the two, as long as the
formation pathway led to enhanced [Na/Fe] and [Al/Fe]. G1's [Na/Fe]
and [Al/Fe] ratios are higher than the value for 47~Tuc; the [Na/Fe]
ratio agrees with previous IL observations of lower mass clusters that
indicate a trend of increasing [Na/Fe] with cluster mass.  This trend
may reflect that the most massive clusters possess larger relative
amounts of Na-enhanced stars.  This result has important consequences
for models that seek to explain the formation of multiple populations
in GCs.

One simple explanation for G1's abundance pattern is that it formed in
a very massive giant molecular cloud, and experienced some amount of
prolonged star formation and self-enrichment.  Under the framework of
\citet{Johnson2017}, the formation of such a massive cluster would
require a very high star formation rate surface density.  Such intense
star formation could be triggered by a major merger in M31's early
assembly, but there does not seem to be any remaining evidence for
such a merger, nor are there other stars or GCs that appear to be
currently associated with G1.

Finally, it is worth noting that despite its unusual properties, G1 is
not unique.  There are several other massive, metal-rich clusters in
M31's halo, some of which may be located in the outer halo.  Any
formation scenario for G1 should also be able to reproduce the
properties of these clusters as well.

This analysis has added to the mystery surrounding G1 and the
formation of M31's outer halo in general
\citep{McConnachie2018,Mackey2019}.  Additional imaging and
spectroscopy of nearby field stars can help assess the presence of
faint streams surrounding G1, while deeper imaging of the cluster
itself could reveal more about possible intracluster abundance
spreads.  Additional IL spectroscopic observations further in the blue
or the infrared could also provide more information about abundance
spreads within the cluster.  On the theoretical front, more detailed
modelling of G1's orbit could reveal associations with other GCs or
stellar streams.  Such information would further help to untangle
M31's complex early assembly, the properties of its satellite dwarfs,
and the nature of massive clusters like G1.

\section*{Acknowledgments}
The authors thank the referee, Claudia Maraston, for suggestions that
have greatly improved this manuscript.
The authors especially thank the observing specialists at Apache Point
Observatory and McDonald Observatory for their assistance with these
observations. The Hobby-Eberly Telescope (HET) is a joint project of
the University of Texas at Austin, the Pennsylvania State University,
Ludwig-Maximilians-Universit\"{a}t M\"{u}nchen, and
Georg-August-Universit\"{a}t G\"{o}ttingen. The HET is named in honor
of its principal benefactors, William P. Hobby and Robert E. Eberly.
GW acknowledges funding from the Kenilworth Fund of the New York
Community Trust.
This work has made use of BaSTI web tools.
This research has made use of the SIMBAD database, operated at CDS,
Strasbourg, France.

\section*{Data Availability}
The data underlying this article will be shared on reasonable
request to the corresponding author after completion of the final
paper in this series.

\footnotesize{

}
\appendix

\section{Abundances from Individual Spectral Lines}\label{appendix:Abunds}

\begin{table*}
\centering
\vspace{-0.25in}
\begin{center}
\caption{Abundances for individual spectral lines.\label{table:LineAbunds}}
  \newcolumntype{d}[1]{D{,}{\;\pm\;}{#1}}
  \newcolumntype{e}[1]{D{.}{.}{#1}}
  \begin{tabular}{@{}llce{2}d{3}d{3}@{}}
    \hline
Element     & Wavelength & EP   & \multicolumn{1}{c}{\cms{$\;\;\log gf$}} & \multicolumn{1}{c}{$\;\;$G1} & \multicolumn{1}{c}{47~Tuc} \\
            & (\AA)      & (eV) &           & \multicolumn{1}{c}{$\;\;\log \epsilon$} & \multicolumn{1}{c}{$\log \epsilon$} \\
    \hline
\ion{Fe}{1}      & 5068.77 & 2.94 & -1.23 & 6.50,0.20 & 6.70,0.10 \\
\ion{Fe}{1}      & 5159.06 & 4.28 & -0.82 & 6.50,0.20 & 6.80,0.10 \\
\ion{Fe}{1}      & 5191.45 & 3.04 & -0.55 & 6.40,0.20 & 6.70,0.10 \\
\ion{Fe}{1}      & 5215.18 & 3.26 & -0.86 & 6.35,0.20 & 6.70,0.10 \\
\ion{Fe}{1}      & 5302.30 & 3.28 & -0.73 & 6.55,0.20 & 6.80,0.10 \\
\ion{Fe}{1}      & 5364.87 & 4.44 &  0.22 & 6.40,0.20 & 6.70,0.10 \\
\ion{Fe}{1}      & 5383.37 & 4.31 &  0.50 & 6.60,0.10 & 6.70,0.10 \\
\ion{Fe}{1}      & 5415.20 & 4.38 &  0.50 & 6.45,0.10 & 6.60,0.10 \\
\ion{Fe}{1}      & 5501.47 & 0.96 & -3.05 & 6.50,0.20 & 6.70,0.10 \\
\ion{Fe}{1}      & 5569.62 & 3.41 & -0.52 & 6.55,0.10 & 6.70,0.10 \\
\ion{Fe}{1}      & 5686.53 & 4.55 & -0.63 & 6.50,0.20 & 6.80,0.10 \\
\ion{Fe}{1}      & 5976.78 & 3.94 & -1.50 & 6.70,0.10 & 6.85,0.10 \\
\ion{Fe}{1}      & 6003.01 & 3.88 & -1.10 & 6.30,0.10 & 6.60,0.10 \\
\ion{Fe}{1}      & 6024.06 & 4.55 & -0.12 & 6.50,0.10 & 6.65,0.10 \\
\ion{Fe}{1}      & 6079.01 & 4.65 & -1.12 & 6.65,0.15 & 6.80,0.05 \\
\ion{Fe}{1}      & 6137.69 & 2.59 & -1.40 & 6.55,0.20 & 6.70,0.20 \\
\ion{Fe}{1}      & 6180.20 & 2.73 & -2.78 & 6.60,0.20 & 6.70,0.10 \\
\ion{Fe}{1}      & 6219.28 & 2.20 & -2.45 & 6.25,0.20 & 6.35,0.10 \\
\ion{Fe}{1}      & 6302.49 & 3.68 & -1.13 & 6.40,0.20 & 6.70,0.10 \\
\ion{Fe}{1}      & 6322.69 & 2.59 & -2.43 & 6.45,0.20 & 6.80,0.10 \\
\ion{Fe}{1}      & 6380.74 & 4.18 & -1.40 & 6.60,0.20 & 6.80,0.10 \\
\ion{Fe}{1}      & 6411.65 & 3.65 & -0.59 & 6.10,0.20 & 6.50,0.10 \\
\ion{Fe}{1}      & 6475.62 & 2.56 & -2.94 & 6.90,0.20 & 7.20,0.10 \\
\ion{Fe}{1}      & 6518.37 & 2.83 & -2.75 & 6.70,0.20 & 7.00,0.10 \\
\ion{Fe}{1}      & 6533.93 & 4.56 & -1.46 & 6.80,0.20 & 6.90,0.10 \\
\ion{Fe}{2}      & 5534.85 & 3.25 & -2.90 & 6.55,0.10 & 6.70,0.10 \\
\ion{Fe}{2}      & 6456.39 & 3.90 & -2.20 & 6.79,0.10 & 6.90,0.10 \\
\ion{Na}{1}      & 5682.63 & 2.10 & -0.70 & 5.78,0.10 & 5.84,0.10 \\
\ion{Na}{1}      & 5688.20 & 2.10 & -0.45 & 5.73,0.10 & 5.84,0.10 \\
\ion{Na}{1}      & 6154.23 & 2.10 & -1.56 & 6.00,0.10 & 5.89,0.10 \\
\ion{Na}{1}      & 6160.75 & 2.10 & -1.26 & 5.92,0.20 & 5.89,0.10 \\
\ion{Mg}{1}      & 5528.31 & 4.34 & -0.62 & 7.00,0.10 & 7.10,0.10 \\
\ion{Mg}{1}      & 5711.09 & 4.34 & -1.83 & 7.00,0.20 & 7.40,0.10 \\
\ion{Al}{1}      & 6696.02 & 3.14 & -1.35 & 6.14,0.20 & 5.95,0.10 \\
\ion{Al}{1}      & 6698.67 & 3.14 & -1.65 & 6.24,0.20 & 6.05,0.10 \\
\ion{Ca}{1}      & 5581.97 & 2.52 & -0.71 & 5.70,0.10 & 5.74,0.10 \\
\ion{Ca}{1}      & 5588.75 & 2.52 &  0.21 & 5.80,0.10 & 5.94,0.10 \\
\ion{Ca}{1}      & 5590.11 & 2.52 & -0.71 & 5.60,0.10 & 5.98,0.10 \\
\ion{Ca}{1}      & 5601.28 & 2.52 & -0.69 & 5.83,0.10 & 5.94,0.10 \\
\ion{Ca}{1}      & 5857.45 & 2.93 &  0.23 & 6.00,0.10 & 6.04,0.10 \\
\ion{Ca}{1}      & 6166.44 & 2.52 & -0.90 & 5.50,0.20 & 5.74,0.10 \\
\ion{Ca}{1}      & 6439.07 & 2.52 &  0.47 & 5.53,0.10 & \multicolumn{1}{c}{$-$} \\
\ion{Ca}{1}      & 6455.60 & 2.52 & -1.35 & 5.75,0.10 & 5.84,0.10 \\
\ion{Ca}{1}      & 6572.78 & 0.00 & -4.29 & 5.73,0.15 & 5.94,0.10 \\
\ion{Ti}{1}      & 4991.07 & 0.84 &  0.45 & 4.24,0.10 & 4.25,0.10 \\
\ion{Ti}{1}      & 4999.50 & 0.83 &  0.32 & 4.34,0.10 & 4.55,0.10 \\
\ion{Ti}{1}      & 5210.38 & 0.05 & -0.82 & 4.44,0.10 & 4.65,0.10 \\
\ion{Ti}{1}      & 6743.12 & 0.90 & -1.63 & 4.24,0.15 & 4.45,0.10 \\
\ion{Ti}{2}      & 5336.79 & 1.58 & -1.60 & 4.24,0.10 & 4.55,0.10 \\
\ion{Ti}{2}      & 5418.77 & 1.58 & -2.13 & 4.24,0.10 & 4.50,0.10 \\
\ion{Cr}{1}      & 5206.02 & 0.94 &  0.02 & 4.53,0.05 & 4.84,0.10 \\
\ion{Cr}{1}      & 5345.80 & 1.00 & -0.95 & 4.53,0.10 & 4.64,0.10 \\
\ion{Mn}{1}$^{a}$ & 4823.52 & 2.32 &  0.14 & 4.42,0.10 & 4.63,0.15 \\
\ion{Mn}{1}$^{a}$ & 5516.77 & 2.18 & -1.85 & 4.32,0.10 & 4.43,0.10 \\
\ion{Mn}{1}$^{a}$ & 6013.51 & 3.07 & -0.35 & 4.12,0.20 & 4.28,0.10 \\
\ion{Ni}{1}      & 6643.63 & 1.68 & -2.22 & 5.21,0.10 & 5.37,0.10 \\
\ion{Ni}{1}      & 6767.77 & 1.83 & -2.14 & 5.31,0.10 & 5.52,0.10 \\
\ion{Cu}{1}$^{a}$ & 5782.13 & 1.64 & -1.72 & 2.58,0.20 & 3.29,0.10 \\
\ion{Zn}{1}      & 4810.53 & 4.08 & -0.14 & 3.85,0.15 & 3.86,0.10 \\
\ion{Y}{2}       & 5087.42 & 1.08 & -0.17 & 1.20,0.20 & 1.46,0.15 \\
\ion{Y}{2}       & 5200.41 & 0.99 & -0.57 & 1.00,0.20 & 1.36,0.10 \\
\ion{Ba}{2}$^{a}$ & 5853.67 & 0.60 & -1.01 & 1.17,0.20 & 1.38,0.20 \\
\ion{Ba}{2}$^{a}$ & 6141.71 & 0.70 & -0.08 & 1.07,0.10 & 1.53,0.05 \\
\ion{Ba}{2}$^{a}$ & 6496.90 & 0.60 & -0.38 & 1.37,0.20 & 1.48,0.10 \\
\ion{Eu}{2}$^{a}$ & 6645.06 & 1.38 &  0.20 & <0.01 & 0.12,0.20 \\
  \hline
\end{tabular}
\end{center}
\medskip
\raggedright $^{a}$ Line has HFS or isotopic splitting.\\
\end{table*}

\end{document}